\documentclass[preprint,showpacs,preprintnumbers,amsmath,amssymb]{revtex4}
\usepackage{amsmath}
\usepackage[normalem]{ulem}
\usepackage{xcolor}
\usepackage{float}
\usepackage{graphicx}
\usepackage[noabbrev]{cleveref}

\begin{document}

\preprint{}
\title{Inclusive jet and dijet productions using $k_t$ and $(z,k_t)\textrm{-factorizations}$ versus ZEUS collaboration data }
\author{\textit{R. Kord Valeshabadi}}
\affiliation {Department of Physics, University of $Tehran$,
1439955961, $Tehran$, Iran.}
\author{\textit{M. Modarres} }
\altaffiliation {Corresponding author, Email:
mmodares@ut.ac.ir,Tel:+98-21-61118645, Fax:+98-21-88004781.}
\author{\textit{S. Rezaie}}
\affiliation {Department of Physics, University of $Tehran$,
1439955961, $Tehran$, Iran.}
\author{\textit{R. Aminzadeh Nik}}
\affiliation {Department of Physics, University of $Tehran$,
    1439955961, $Tehran$, Iran.}
\date{\today}
\begin{abstract}
        In this paper, we investigate the  differential cross sections of the inclusive jet
        and dijet productions of the ZEUS collaboration data  at the center
         of mass energies of $\sim 300\;GeV$ and $ 319\;GeV$ using the $k_t$ and $(z,k_t)\textrm{-factorizations}$
         with the different unintegrated and double unintegrated parton distribution functions, i.e., UPDFs and DUPDFs,
         respectively. The \textsc{KaTie} event generator is used to calculate the differential
         cross section with the UPDFs, while for the input DUPDFs the calculations are directly
         performed by evaluating the corresponding matrix elements. We check the effect of
         choosing the different implementation of angular or strong ordering constraints
         using the UPDFs and the corresponding DUPDFs of Kimber-Martin-Ryskin (KMR) and the
         leading-order (LO) and next-to-leading-order (NLO) Martin-Ryskin-Watt (MRW) approaches.
         The impacts of choosing virtualities $k^2 = k_t^2$ or $k^2 = {k_t^2\over {(1-z)}}$ in the differential cross
         section predictions for  the ZEUS experimental data are also investigated.  It is observed that, as one should expect, 
         the applications of $(z,k_t)\textrm{-factorization}$ is better than the $k_t\textrm{-factorization}$ framework
         for the predictions of high virtuality $Q^2$ with respect to the ZEUS collaboration data, and also the results
          of the  KMR and LO-MRW UPDFs and DUPDFs    are reasonably close to each other and in general can describe the data. It is also observed that only in the case of  $k_t$-factorization,    the inclusion of  Born level makes our results to overshoot the  inclusive jet experimental data.
\end{abstract}
\maketitle
\section{Introduction}
   The  precise theoretical prediction of hadron level differential cross sections is a crucial issue since the advent
    of collider experiments. Two conventional methods exist for calculating theoretical cross sections. The first framework which is more common and widely used is the so  called the collinear
    factorization method \cite{collinear-factorization,collins_QCD} and the second one that is attracted more interests in the recent years is the $k_t\textrm{-factorization}$ formalism ($k_t$ is the transverse momentum of parton) \cite{kt_factorization1,kt_factorization2}, for good reviews see \cite{smallXReview1,smallXReview2,smallXReview3}. But recently the $k_t\textrm{-factorization}$ approach is extended and generalized and led to the presentation of  the $(z, k_t)\textrm{-factorization}$ formalism ($z$ is the longitudinal fractional momentum of parton) \cite{wattDoubleJet}.

 The collinear factorization theorem allows us to write the hadronic cross section as a convolution of
 the partonic cross section and the parton distribution functions (PDFs). It is assumed that the parton
 enters into the hard scattering is collinear with the parent hadron and hence the PDFs do not depend on
 the transverse momentum of this parton. This framework which is well established  becomes a conventional
 approach and  used for the cross section calculations of many processes up to the next-to-next-to-leading-order
 (NNLO) \cite{nnlo_jet,2gnnlo,wjetnnlo,wpwmnnlo,ttbarnnlo} and the next-to-next-to-next-to-leading-order
 (NNNLO) \cite{n3lo_jet,nnnlo-higgs} in the QCD perturbation theory. Obtaining the cross sections to
 such a high order of perturbation theory requires the use of novel and cumbersome techniques for calculating
 the matrix elements \cite{multi_loop} and curing the infrared divergences \cite{Njettiness,qtsubtraction,proj_to_born}.
 Apart from these difficulties, it is also  very computationally intensive.

On the other hand, the $k_t\textrm{-factorization}$ framework is the extension of  the
collinear-factorization framework and considers the incoming
parton which enters into the hard interaction also to have
transverse component orthogonal to the direction of parent hadron.
Therefore working in this framework requires the unintegrated
parton distribution functions (UPDFs) which are transverse
momentum dependent. The UPDFs are indispensable part of the
calculation and they play a vital role in the numerical
predictions of cross sections. Currently, different methods are
proposed for the UPDFs and no unique one exists in which all
researchers are agreed upon it. Among these methods the
Kimber-Martin-Ryskin (KMR) \cite{KMR} and  the Martin-Ryskin-Watt at
leading order (LO-MRW) \cite{MRW} and  
next-to-leading order (NLO-MRW) UPDFs \cite{MRW} are thoroughly
investigated in the references \cite{Mod1,Mod2,Mod3,Mod4,Mod5,
Mod6,Mod7} and shown a great success since their presentations
\cite{Mod8,Mod9,Mod10,Mod11,Mod12,Mod13,Mod14,Mod15,Mod16,Lipatov_photon,Baranov_Drell,lipatov_jet}.

Working in the above frameworks, have this advantage  that with   some lower order Feynman diagrams, one can obtain a  comparable precision with respect to 
the same cross section calculation, but   with the collinear factorization framework in higher order.   Despite these
  success,   the $k_t$-factorizations are only limited to
high energy  domain \cite{wattDoubleJet} and the small $x$ region,
($x$ is the longitudinal fractional momentum of parton or the
Bjorken scale).

One way to extend and generalize the $k_t\textrm{-factorization}$
framework is proposed by Watt et al,
\cite{DUPDFElectroweak},  which is called the $(z,
k_t)\textrm{-factorization}$ method. In this approach, they
extend the partonic cross section to one step before the hard
scattering, and instead of UPDFs, they used the double unintegrated
parton distribution functions (DUPDFs) which are generated by
LO-MRW UPDFs. This framework was tested for calculating the
differential cross sections of inclusive jet \cite{wattDoubleJet}
and electroweak bosons productions~\cite{DUPDFElectroweak}, which
was remarkably successful in explaining of the data. Also, it is
worth to mention that one of the advantage of the
$(z,k_t)\textrm{-factorization}$ formalism  with respect to the
collinear and the $k_t\textrm{-factorization}$ approaches is that,
with a calculation of lower order sub-processes in the perturbation
theory one can achieve acceptable results.

The KMR, LO-MRW and NLO-MRW UPDFs can be obtained from two
alternative formulas, i.e.,the  integral and differential definitions.
At first, it seems that the integral form of the UPDFs can be
obtained from the differential form. But in the reference
\cite{Golec_on_KMR}, it is  shown that these two different
approaches     give   different results at the
virtuality larger than the factorization scale, $\mu$. It was
demonstrate that for obtaining correct results, one should either
use the differential form with the cutoff dependent PDFs, which is
troublesome and inefficient, or the integral form, using the PDFs
from the global fit, which is efficient and often used in the
literature. There is also a debate over the choice of the cutoff
introduced in these UPDFs, i.e., the cutoff based upon the angular
or strong ordering constraints. The angular ordering cutoff allows
the parton to have transverse momentum larger than $\mu$, while
strong ordering cutoff limits the transverse momentum to the
region less than $\mu$. In the reference \cite{Guiot_heavy_quark}
it was claimed that in the heavy quark production the angular
ordering cutoff leads to the result which considerably
overestimates the data. But, it can be realized from some parts of
this report  \cite{Guiot_heavy_quark} that, it
is erroneously used the wrong form of the UPDFs formula in the
$k_t > \mu$ region, by claiming that the Sudakov form factor is
greater than one for $k_t
> \mu$, which is physically incorrect and should be set equal to
one,  as it can be seen in the references
\cite{KMR,MRW}. However, in the reference \cite{Guiot_pathology}
the form of the UPDFs formula is corrected and the Sudakov form
factor is set equal to one ($k_t > \mu$), but still it is insisted
that the UPDFs with angular ordering constraint greatly
overestimates the data, without referring to this fact that the wrong UPDFs
definitions are used in the reference \cite{Guiot_heavy_quark}. However, in the
present work it will  be shown that, at least in the case of inclusive
dijet production the UPDFs with angular ordering constraint does
not have as significant impact as what is mentioned in the
references \cite{Guiot_heavy_quark,Guiot_pathology}, and the
calculated cross sections are similar to the strong ordering case.
It is also worth to point out that, recently another approach for
constructing transverse momentum dependent parton distribution
functions, i.e., TMD-PDFs, led its way into the
$k_t\textrm{-factorization}$ framework from the parton branching (PB),
i.e., solution of the QCD evolution equation \cite{PB1,PB2}. In
the reference \cite{PB_dynamical_res}, the LO-MRW UPDFs and the
TMD-PDFs obtained from the PB are compared to each other and shown
that these two distributions are close to each other at middle
transverse momentum, but are different from each other at the
extremely large $k_t \gg \mu$ and small $k_t \ll \mu$.
 The comparison of the UPDFs of LO-MRW and PB is also
shown in the Drell-Yan Z-boson $p_T^{ll}$ distribution, in which
these two distributions give the different results at large and
small $p_{T}^{ll}$, such that those of PB are closer to the data
\cite{PB_dynamical_res}.
 
In this study, we first give a brief review of the $k_t$  and  
$(z, k_t)\textrm{-factorizations}$ approaches, related to the KMR, LO-MRW and NLO-MRW UPDFS and DUPDFs,  in the sections
(\ref{kt-factorisation}) and (\ref{zkt}), respectively.   We present a
comparison between the experimental data of inclusive jet
\cite{zeus2002} and dijet differential cross sections \cite{Zeus_2010} of ZEUS
collaboration data  at the center of mass energies of $\sim
300\;GeV$ and $319\;GeV$, respectively, and theoretical
predictions using the $k_t$ and the $(z,
k_t)\textrm{-factorizations}$ with their corresponding
aforementioned UPDFs and DUPDFs to test the credibility and
precision of the KMR and MRW frameworks (the section
\ref{Numerical-Results}). Additionally we also check the variations
of each model versus the corresponding collinear framework \cite{nnlo_dijet,zeus2002} by
calculating their ratios with respect to the data.
\section{ The  $k_t$-factorization theorem}\label{kt-factorisation} 
The $k_t\textrm{-factorization}$ approach allows us to write the
electron-proton cross section as a convolution of UPDFs, which
have the phenomenological origin because of the PDFs input, and
the process dependent parton level cross section.
 In this framework, the transverse momentum, $k_t$, of the parton coming into the hard process
 is not neglected, unlike the collinear factorization. This formalism mostly becomes important
 in the limit of small $x$, or equivalently where the transverse momentum of parton coming
 into hard interaction becomes comparable against the collinear part of momentum.
Using the $k_t\textrm{-factorization}$ framework allows us to
write the cross section for DIS process as \cite{wattDoubleJet}:
\begin{equation}
\label{eq:one}
    \sigma_{T,L}^{\gamma^*p} = \sum_a \int_0^1\;\frac{\mathrm{d}x}{x}\int_0^\infty\frac{\mathrm{d}k_t^2}{k_t^2}f_{a}(x,k_t^2,\mu^2)
     \hat{\sigma}^{\gamma^*a}(x,k_t^2,\mu^2).
\end{equation}
In this formula, $f_{a}$ is the partonic UPDF ($a=q, \bar q$ and $g$) and
$\hat{\sigma}^{\gamma^*a}$ is the partonic cross section of
virtual photon with the corresponding parton, which can be calculated with the
perturbative QCD.

The two essential parts of the above \cref{eq:one} are obtaining a
suitable UPDFs and calculating the off-shell partonic cross
sections. In the present report the UPDFs are generated using the
KMR, LO-MRW and NLO-MRW approaches. For calculating the cross
sections, the \textsc{KaTie}  parton-level event generator \cite{KATIE} is used,
which automatically takes care of the off-shell partonic cross
sections and gives the hadronic cross sections with desirable
accuracy. This Monte Carlo event generator is discussed after a
short review of the UPDFs models.
\subsection{The UPDFs}\label{UPDFs}
The UPDFs of the proton, $f_{a}(x,k_t^2,\mu^2)$, is
defined~\cite{KimberThesis} as the probability of finding a parton
inside the proton with the longitudinal momentum fraction $x$ and
the transverse momentum $k_t$ at scale $\mu$; It also should
satisfy normalization condition:
\begin{equation}
\label{eq:two}
    a(x,\mu^2) = \int_0^{\mu^2}\;\frac{\mathrm{d} k_t^2}{k_t^2}
    f_a(x,k_t^2,\mu^2),
\end{equation}
where $a(x,\mu^2)$ in the above equation is the collinear PDF.

The  KMR and MRW approaches suggested double scale, $k_t^2$ and $\mu^2$, dependent PDFs,
 $f_a(x, k_t^2,\mu^2)$ based on the Dokshitzer-Gribov-Lipatov-Altarelli-Parisi (DGLAP)
 evolution equations \cite{gribovDeepInelastic, altarelli,DokshitzerDeepInelastic},
 which satisfies the normalization condition and can easily be calculated numerically for (anti) quarks and gluons.
\subsubsection{The KMR UPDFs prescription}\label{KMR-Formalism}
In the KMR formalism, it is assumed that the transverse momentum
of the  parton along the evolution ladder is strongly ordered up
to the final evolution step. In the last step this assumption
breaks down and the incoming parton enters into the hard
interaction posses the large transverse momentum ($k_t \simeq \mu
$). Eventually, the $k_t$-dependent parton evolves to the scale
$\mu$ via the Sudakov form factors, without any real emission, and
therefore the parton transverse momentum stays unchanged.
Generally, this model can be formulated as:
    \begin{equation}\label{eq:three}
        f_a(x, k_t^2, \mu^2) = T_a(k_t^2,\mu^2)\left(\frac{\alpha_{s}(k_t^2)}{2 \pi}  \sum_{a^\prime=q,g}
         \int_x^{1-\Delta} P_{aa^\prime}(z)\frac{x}{z} a^\prime(\frac{x}{z}, k_t^2)\; \mathrm{d}z \right),
    \end{equation}
where $T_a(k_t^2,\mu^2)$ is the \textit{Sudakov Form Factor} which
resums all the virtual contributions from the scale $k_t$ to the
scale $\mu$:
    \begin{equation}\label{eq:four}
    \begin{split}
    T_a(k_t^2 < \mu^2,\mu^2) &= exp\left(-\int_{k_t^2}^{\mu^2}\frac{\mathrm{d}\kappa_t^2}{\kappa_t^2} \frac{\alpha_s(\kappa_t^2)}{2 \pi}\;
     \sum_{a^\prime}\int_{0}^{1-\Delta} P_{a^\prime a}(\xi)\;\mathrm{d}\xi \right),
        \\
    T_a(k_t^2 \geq \mu^2 ,\mu^2) &= 1.
    \end{split}
    \end{equation}
    In the equations (\ref{eq:three}) and (\ref{eq:four}) $P_{aa^\prime}$ and $P_{a^\prime a}$ are
    the unregularized splitting functions, which for $P_{qq}$ and $P_{gg}$ are divergent in the
     limit $z \to 1$, i.e. the soft gluon emission region. To circumvent such a divergence,
     the upper limit of the integrals over $z$ in the equations (\ref{eq:three}) and (\ref{eq:four})
     are changed from $1$ to $1-\Delta$. However in the above framework this cutoff is imposed on
     the soft (anti) quark emissions in addition to the soft gluon cases, which is theoretically
     incorrect. To determine $\Delta$, the KMR formalism uses the advantage of angular ordering
     of the gluons emissions \cite{AngularOrdering} in the last step of the evolution ladder, and obtain it as \cite{KimberThesis,KMR}:
\begin{equation}\label{eq:five}
    \Delta = \frac{k_t}{k_t + \mu}.
\end{equation}

However, it is also standard to adopt another choice of  cutoff,
according to the strong ordering in the transverse momentum of the
emitted partons, which resulted as \cite{kimber_SO}:
\begin{equation}
\label{eq:six}
    \Delta = \frac{k_t}{\mu}.
\end{equation}
This limits the UPDFs to the region of $k_t < \mu$, while in the
cutoff derived from the angular ordering the transverse momentum
of the parton is free to exceed $\mu$ \cite{Golec_on_KMR}.
 Also the unitarity condition gives the same cutoff for the
virtual part \cite{wattDoubleJet}, i.e. \cref{eq:five,eq:six}, but
by replacing $k_t$ with $\kappa_t$. In the section
\ref{Numerical-Results},  the application of LO-MRW UPDF model
with the strong and angular ordering cutoffs on the
$k_t\textrm{-factorization}$ prediction of the high $Q^2$
inclusive jet and dijet data is shown.

In general one should note that in both of the KMR and MRW
approaches are restricted to the $k_t \ge \mu_0$ limit, where
$\mu_0 \approx  1 \; GeV$. Therefore for the $k_t < \mu_0$ we use
the follwing form  mentioned in the reference \cite{KMR,MRW},
where therein the authors assume the density of partons are
constant at fixed $x$ and $\mu^2$ and also must satisfy the
normalization condition the \cref{eq:two}:
\begin{equation} \label{eq:seven}
 \frac{1}{k_t^2} f_a(x, k_t^2 <\mu_0^2, \mu^2) = \frac{1}{\mu_0^2}a(x,\mu_0^2)T_a(\mu_0^2,\mu^2).
\end{equation}

Finally we can write the KMR UPDFs and their corresponding Sudakov
form factors for (anti) quarks and gluons, respectively, as:

\begin{equation}\label{eq:eight}
\begin{split}
f_{q}^{\textrm{KMR}}(x,k_t^2, \mu^2) = T_{q}^{\textrm{KMR}}(k_t^2,
\mu^2) \frac{\alpha_{s}(k_t^2)}{2\pi}\int_x^{1-\Delta} \big[
P_{qq}^{LO}(z)\frac{x}{z}q^{LO}(\frac{x}{z}, k_t^2) +\\
P_{qg}^{LO}(z) \frac{x}{z} g^{LO}(\frac{x}{z},
k_t^2)\big]\;\mathrm{d}z,
\end{split}
\end{equation}

\begin{equation}\label{eq:nine}
T_q^{\textrm{KMR}}(k_t^2,\mu^2) = exp\left(-
\int^{\mu^2}_{k_t^2}\frac{\mathrm{d}\kappa_t^2}{\kappa_t^2}
\frac{\alpha_{s}(\kappa_t^2)}{2 \pi} \int_{0}^{1-\Delta}
P_{qq}^{LO}(\xi)\;\mathrm{d}\xi \right),
\end{equation}

and
\begin{equation}\label{eq:ten}
\begin{split}
f_g^{\textrm{KMR}}(x,k_t^2, \mu^2) = T_g^{\textrm{KMR}}(k_t^2, \mu^2)
\frac{\alpha_{s}(k_t^2)}{2\pi}\int_x^{1-\Delta}\big[
P_{gg}^{LO}(z)\frac{x}{z}g^{LO}(\frac{x}{z}, k_t^2) +\\ \sum_q
P_{gq}^{LO}(z) \frac{x}{z} q^{LO}(\frac{x}{z},
k_t^2)\big]\;\mathrm{d}z,
\end{split}
\end{equation}

\begin{equation}\label{eq:eleven}
T_g^{\textrm{KMR}}(k_t^2,\mu^2) = exp\left(-\int_{k_{t}^2}^{\mu^2}
\frac{\mathrm{d}\kappa_t^2}{\kappa_t^2}
\frac{\alpha_{s}(\kappa_{t}^{2})}{2 \pi}\int_{0}^{1-\Delta} (\xi
P_{gg}^{LO}(\xi) + n_f P_{qg}^{LO}(\xi))\;\mathrm{d}\xi \right).
\end{equation}
where $q={u,\overline{u}, d, \overline{d},\ldots}$.

\subsubsection{The MRW UPDFs prescription at LO and NLO levels} \label{MRW-Formalism}
Martin, et al \cite{MRW} corrected the imposition of the cutoff on
the soft (anti) quark emissions in the KMR prescription and
changed the aforementioned KMR UPDFs with the angular ordering
constraint such that to be only applied on the soft gluon
emissions. As a result the LO-MRW formalism takes the following
forms for the (anti) quarks and the gluons, respectively, as
follows:
\begin{equation}\label{eq:twelve}
\begin{split}
f_q^{\textrm{LO-MRW}}(x,k_t^2, \mu^2) = T_q^{\textrm{LO-MRW}}(k_t^2, \mu^2)
 \frac{\alpha_{s}(k_t^2)}{2\pi}\int_x^1\Big[ P_{qq}^{LO}(z)\frac{x}{z}q^{LO}(\frac{x}{z}, k_t^2)\Theta(1- z-\Delta)  \\
+ P_{qg}^{LO}(z) \frac{x}{z} g^{LO}(\frac{x}{z},
k_t^{2})\Big]\;\mathrm{d}z,
\end{split}
\end{equation}
\begin{equation}\label{eq:thirteen}
T_q^{\textrm{LO-MRW}}(k_t^2,\mu^2) = exp\left(-\int_{k_t^2}^{\mu^2}
\frac{\mathrm{d}\kappa_t^2}{\kappa_t^2}
\frac{\alpha_{s}(\kappa_t^2)}{2 \pi}\int_{0}^1
P_{qq}^{LO}(\xi)\Theta(1 - \xi -\Delta)\;\mathrm{d}\xi \right),
\end{equation}
and
\begin{equation}\label{eq:fourteen}
\begin{split}
f_g^{\textrm{LO-MRW}}(x,k_t^2, \mu^2) = T_g^{\textrm{LO-MRW}}(k_t^2, \mu^2) \frac{\alpha_{s}(k_t^2)}{2\pi}\int_x^1\Big[  P_{gg}^{LO}(z)\frac{x}{z}g^{LO}(\frac{x}{z}, k_t^2)\Theta(1 - z-\Delta)\\
+ \sum_q P_{gq}^{LO}(z)\frac{x}{z} q^{LO}(\frac{x}{z},
k_t^2)\Big]\;\mathrm{d}z,
\end{split}
\end{equation}
\begin{equation}\label{eq:fifteen}
\begin{split}
T_g^{\textrm{LO-MRW}}(k_t^2,\mu^2) = exp\Big(-\int_{k_t^{2}}^{\mu^2}\frac{\mathrm{d}\kappa_t^2}{\kappa_t^2} \frac{\alpha_{s}(\kappa_t^2)}{2 \pi}\int_{0}^1 ( \xi P_{gg}^{LO}(\xi)\Theta(1 - \xi-\Delta)\Theta(\xi-\Delta) \\
 + n_f P_{qg}^{LO}(\xi))\;\mathrm{d}\xi \Big).
 \end{split}
\end{equation}
where $q={u,\overline{u}, d, \overline{d},\ldots}$.

They also extended the LO-MRW formalism to the NLO level (NLO-MRW)
\cite{MRW}, which follows from the NLO DGLAP evolution equation.
At this order the virtuality takes the following form instead of
$k_t^2$:
\begin{equation} \label{eq:sixteen}
    k^2 = \frac{k_t^2}{1-z}.
\end{equation}
We can write the quark and the gluon UPDFs at the NLO level
according to \cite{MRW}:
\begin{equation}\label{eq:seventeen}
\begin{split}
    f_q^{\textrm{NLO-MRW}}(x,k_t^2,\mu^2)= \int_{x}^{1}\mathrm{d}z\;T_q^{\textrm{NLO-MRW}}(k^2,\mu^2)\;\frac{{\alpha_{s}(k^2)}}{2\pi} \bigg[ \tilde{P}_{qq}^{(0+1)}(z)\frac{x}{z}q^{NLO}(\frac{x}{z}, k^2) \\  + \tilde{P}_{qg}^{(0+1)}(z) \frac{x}{z} g^{NLO}(\frac{x}{z}, k^2)\bigg]\;\Theta(1-z-k_t^2/\mu^2),
\end{split}
\end{equation}
\begin{equation}\label{eq:eighteen}
    T_q^{\textrm{NLO-MRW}}(k^2,\mu^2) = exp\left(-\int_{k^2}^{\mu^2}  \frac{\mathrm{d}\kappa^2}{\kappa^2}
    \frac{\alpha_{s}(\kappa^2)}{2 \pi}\; \int_{0}^{1} \;\mathrm{d}\xi \; \xi[\tilde{P}_{qq}^{(0+1)}(\xi) + \tilde{P}_{gq}^{(0+1)}(\xi)] \right),
\end{equation}
and:
\begin{equation}\label{eq:nineteen}
\begin{split}
    f_g^{\textrm{NLO-MRW}}(x,k_t^2,\mu^2)= \int_{x}^{1}\mathrm{d}z\;T_g^{\textrm{NLO-MRW}}(k^2,\mu^2)
    \;\frac{{\alpha_{s}(k^2)}}{2\pi}\bigg[ \tilde{P}_{gg}^{(0+1)}(z)\frac{x}{z}g^{NLO}(\frac{x}{z}, k^2) \\
    + \sum_q \tilde{P}_{gq}^{(0+1)}(z) \frac{x}{z} q^{NLO}(\frac{x}{z}, k^2)\bigg]\;\Theta(1-z-k_t^2/\mu^2),
\end{split}
\end{equation}
\begin{equation}\label{eq:twenty}
    T_g^{\textrm{NLO-MRW}}(k^2,\mu^2) = exp\left(-\int_{k^2}^{\mu^2}  \frac{\mathrm{d}\kappa^2}{\kappa^2}
     \frac{\alpha_{s}(\kappa^2)}{2 \pi}\; \int_{0}^{1} \;\mathrm{d}\xi \; \xi \left[\tilde{P}_{gg}^{(0+1)}(\xi)+2n_F \tilde{P}_{qg}^{(0+1)}(\xi)\right] \right).
\end{equation}
where $q={u,\overline{u}, d, \overline{d},\ldots}$, and
$\tilde{P}_{ab}^{(0+1)}$s in above equations, where $a$, $b$ can
be $q$ or $g$, are defined in the reference \cite{MRW}. It is also
seen in the above formula  that an additional Heaviside step
function is imposed on the UPDFs, which stops the parton to have
virtuality larger than factorization scale, i.e. $\Theta(\mu^2 -
k^2) = \Theta(1-z-k_t^2/\mu^2)$. In the above equations
$\kappa^2={\kappa_t^2\over {1-\xi}}$.

Finally, for the UPDFs of MRW approach in the $k_t < \mu_0 \approx
1\;GeV$, at the LO and the NLO levels we again use the similar
\cref{eq:seven}, with this difference that the corresponding PDFs
and the Sudakov form factors are used.

\subsection{The $k_t$-factorization approach in the inclusive jet and dijet differential cross sections}\label{KATIE_N}

In this work, for the cross section calculation of high-$Q^2$
inclusive jet \cite{zeus2002} and dijet \cite{Zeus_2010}
productions in the $k_t\textrm{-factorization}$ framework, we
utilize the \textsc{KaTie} event generator \cite{KATIE}. Providing
the input UPDFs, this generator can produce the differential cross
sections in the $k_t\textrm{-factorization}$ framework with the
desirable accuracy while saving the gauge invariance.

One important point about using this event generator is the
difference between the definitions of cross section formula in the
\cref{eq:one} and the normalization condition in the \cref{eq:two}
with the corresponding formulas in the reference \cite{KATIE}. By
looking at the equations (2) and (3) of the reference
\cite{KATIE}, one notices that the following normalization
condition is used:
\begin{equation}
\label{eq:twenty_one} a(x,\mu^2) = \int_0^{\mu^2}\mathrm{d} k_t^2
f_a(x,k_t^2,\mu^2).
\end{equation}
This means that the UPDFs definitions of the reference
\cite{KATIE} have the $GeV^{-2}$ dimension while the \cref{eq:two}
which we employed in the section \ref{UPDFs}, follows from the
original KMR and MRW definitions \cite{KMR,MRW}, which are
dimensionless. Also, the cross section formula, i.e., the equation
(1) of the reference \cite{KATIE}, does not have the virtuality $1
/ k_t^2$. In other word, $1 / k_t^2$ in the cross section of the
reference \cite{KATIE} is moved from the cross section formula
into the UPDFs definition. In view of this fact, we can change
UPDFs of KMR, LO-MRW and NLO-MRW in the section \ref{UPDFs},
$f_a(x, k_t^2, \mu^2)$, to $f_a(x, k_t^2, \mu^2) / k_t^2$, in
order to be consistent with the \textsc{KaTie} event generator and
obtain correct results.

Before moving forward about how to use this generator, an
important remark about the current version of this generator is in
order. In generating the results of inclusive jet and dijet
experiments of the ZEUS collaboration data within the
$k_t$-factorization framework, we noticed that our results
significantly offshoot the data that leaded us to doubt about
results. So, we decided to run this generator in the collinear
mode, which again the outputs were strangely offshoot the data.
Therefore, it becomes  clear that something is going wrong, and we
diagnosed that the lab to Breit transformation is not
appropriately implemented in KaTie. In fact, in the collinear
factorization framework the parton should be scattered back along
the $-\hat{z}$ direction for the Born level subprocess
$\gamma^\ast  + q \to q$, without having any transverse momentum
which did not happen in KaTie  \cite{KATIE}. Therefore, we fixed
this lab to Breit frame transformation \cite{KATIE1} according to the matrix
elements of   the reference
\cite{Devenish:2004pb}.

To work with this generator, the UPDFs should be provided as a
grid file of four columns of $ln(x)$, $ln(|k_t|^2)$, $ln(\mu^2)$
and $f(x,|k_t|,\mu)$. Due to the consistency of this generator
with TMDLIB \cite{TMDLIB}, we generate the UPDFs in accordance
with the UPDFs grid files of this library. For each aforementioned
UPDF models, i.e. KMR and MRW, the total of the eleven files
including the gluon, the first five flavors of quarks and their
corresponding anti-quarks are generated. LO-PDFs in the
\cref{eq:eight,eq:ten,eq:twelve,eq:fourteen} and the NLO-PDFs in
the \cref{eq:seventeen,eq:nineteen} are utilized correspondingly
with the central PDFs of MMHT2014lo68cl and MMHT2014nlo68cl
\cite{harland-lang_uncertainties_2015,harland-lang_charm_2016} of
LHAPDF6 library~\cite{LHAPDF6}. Finally, for obtaining the UPDFs
integrals we resort to the Gauss-Legendre quadrature method of the
BOOST C++ libraries~\cite{BOOST_CPP}.

After preparing the UPDFs, we can follow the \textsc{KaTie} event
generator to set the desired partonic sub-processes, the
renormalization and factorization scales, and the experimental
cuts. In this project   $\gamma^\ast+q \to g + q$ and
$\gamma^\ast+g\to q + \bar{q}$ subprocesses,   see the figure $2$,
are included in the numerical calculation and also the
renormalization and factorization scales $\mu_{R,F}^2 = Q^2$ are
adopted.  Finally the experimental cuts are set according to the
high $Q^2$ inclusive jet and dijet productions in the $ep$
scattering of the aforementioned ZEUS collaboration experiments,
where in the section \ref{Numerical-Results} the necessary
information about them are mentioned.
\section{ The   $(z, k_t)$-factorization formalism }\label{zkt}
Watt, et al \cite{wattDoubleJet} generalized the
$k_t$-factorization approach to the $(z,k_t)$-factorization
formalism. With this generalization, the last step emission   along
the evolution ladder also participates into the processes, and
as a result, leads to a improved prediction of the cross
sections, especially at large $z$. 

In this formalism, the virtuality has the form of the
\cref{eq:sixteen}, $k^2 = k_t^2 / (1-z)$, in contrast to $k_t^2$
in case of the $k_t\textrm{-factorization}$, which holds in the high
energy limit, $z \rightarrow 0$. Such  a selection of
virtuality has this effect that the cross section becomes no more limited on the
parton enters into the hard process, and the parton emitted in the
last step, plays an important role in the cross section calculations. It is worth to mention that, although in the NLO-MRW formalism, the same
selection of virtuality is used but it does not have the features
of the $(z,k_t)\textrm{-factorization}$ framework, because the
dependency on $z$ is integrated out.

The cross section for $\gamma^{\ast}p$ in the $(z,
k_t)$-factorization, can be generally expressed as
\cite{wattDoubleJet}:
\begin{equation}
\label{eq:twenty_two}
    \sigma_{T,L}^{\gamma^*p} = \sum_a \int_0^1\;\frac{\mathrm{d}x}{x}\int_x^1\frac{\mathrm{d}z}{z}
    \int_0^\infty\frac{\mathrm{d}k_t^2}{k_t^2}f_{a}(x, z, k_t^2,\mu^2) \hat{\sigma}_{T,L}^{\gamma^*a}(x, z, k_t^2,\mu^2),
\end{equation}
where in this equation, $f_{a}(x, z, k_t^2,\mu^2)$ is called
DUPDFs, and $\hat{\sigma}_{T,L}^{\gamma^*a}$ is the transverse and
longitudinal components of partonic cross section which now
depends also on $z$ in addition to   $k_t^2$ and   $\mu^2$.
Hence for the inclusive jet and dijet cross sections, we need to
obtain both the DUPDFs and the partonic cross sections. In the
following we are going to discuss briefly the methods of
obtaining DUPDFs and also calculating the partonic cross sections.
\subsection{The DUPDFs}\label{DUPDF}
The DUPDFs which are unintegrated over both $k_t$ and $z$ can be
obtained according to the reference \cite{wattDoubleJet} via
exactly the same way as we did for the UPDFs and mentioned before
in the section \ref{UPDFs}, provided that the integration over $z$ is
ignored. The DUPDFs which are introduced in the reference
\cite{wattDoubleJet} is based only on the LO-MRW approach,
although it seems to be working well and   successfully
can describe the data \cite{wattDoubleJet, DUPDFElectroweak}. But at
the same time, it seems one of the necessary component is
overlooked and as can be seen in the LO-MRW UPDFs, i.e., the
\cref{eq:twelve,eq:thirteen,eq:fourteen,eq:fifteen}, that the
scale $k_t^2$ is used in the arguments of both the input PDFs and
the coupling constant, while the expectation is that the scale
should be $k_t^2/(1-z)$, the \cref{eq:sixteen}. It means that the
correct form for the DUPDFs should follow from the
\cref{eq:seventeen,eq:eighteen,eq:nineteen,eq:twenty}. Here in
this paper, for investigation of the role virtuality, three
different forms of DUPDFs based upon the UPDFs approaches of the
references \cite{KMR,wattDoubleJet} and also \cite{MRW} are
generated.
\subsubsection{The KMR DUPDFs prescription}\label{KMRDUPDF}
The first approach that is based on the KMR prescription, in which the cutoff
on $z\rightarrow 1$ is imposed over the last step soft (anti)
quark and the gluon emissions, gives the DUPDFs (it is referred to
DKMR in our results) for the (anti) quark and gluon contributions
as, respectively:
\begin{equation}
\begin{split}
f_q^{\textrm{DKMR}}(x,z, k_t^2, \mu^2) = T_q^{\textrm{DKMR}}(k_t^2,
\mu^2) \frac{\alpha_{s}(k_t^2)}{2\pi} \bigg[
P_{qq}^{LO}(z)\frac{x}{z}q^{LO}(\frac{x}{z}, k_t^2) \Theta(1-
z-\Delta)\\ + P_{qg}^{LO}(z) \frac{x}{z} g^{LO}(\frac{x}{z},
k_t^2)\Theta(1- z-\Delta) \bigg],
\end{split}
\end{equation}
\begin{equation}
T_q^{\textrm{DKMR}}(k_t^2,\mu^2) = exp\left(- \int^{\mu^2}_{k_t^2}
\frac{\mathrm{d}\kappa_t^2}{\kappa_t^2}
\frac{\alpha_{s}(\kappa_t^2)}{2 \pi}\int_{0}^{1-\Delta}
P_{qq}^{LO}(\xi) \;\mathrm{d}\xi \right),
\end{equation}
\begin{equation}
\begin{split}
f_g^{\textrm{DKMR}}(x,z,k_t^2, \mu^2) = T_g(k_t^2, \mu^2)
\frac{\alpha_{s}(k_t^2)}{2\pi}\bigg[
P^{LO}_{gg}(z)\frac{x}{z}g^{LO}(\frac{x}{z}, k_t^2)\Theta(1- z-\Delta)
\\ + \sum_q P_{gq}^{LO}(z) \frac{x}{z} q^{LO}(\frac{x}{z},
k_t^2)\Theta(1- z-\Delta) \bigg],
\end{split}
\end{equation}
\begin{equation}
T_g^{\textrm{DKMR}}(k_t^2,\mu^2) = exp\left(-\int_{k_{t}^2}^{\mu^2}
\frac{\mathrm{d}\kappa_t^2}{\kappa_t^2}
\frac{\alpha_{s}(\kappa_{t}^{2})}{2 \pi} \int_{0}^{1-\Delta} (\xi
P_{gg}^{LO}(\xi) + n_f P_{qg}^{LO}(\xi))\;\mathrm{d}\xi \right),
\end{equation}
where $q={u,\overline{u}, d, \overline{d},\ldots}$.
\subsubsection{The MRW DUPDFs prescriptions at LO level}\label{MRWDPDUF}
The second method which is based on the MRW approach, in which the cutoff is
only applied to the soft gluon emissions, gives the following
DUPDFs (it is referred to DMRW in our results) for the (anti)
quark and the gluon, respectively:
\begin{equation}
\begin{split}
f_q^{\textrm{DMRW}}(x, z, k_t^2, \mu^2) = T_q^{\textrm{DMRW}}(k_t^2,
\mu^2) \frac{\alpha_{s}(k_t^2)}{2\pi}\bigg[
P_{qq}^{LO}(z)\frac{x}{z}q^{LO}(\frac{x}{z}, k_t^2)\Theta(1- z-\Delta)
\\ + P_{qg}^{LO}(z) \frac{x}{z} g^{LO}(\frac{x}{z},
k_t^{2})\bigg],
\end{split}
\end{equation}
\begin{equation}
T_q^{\textrm{DMRW}}(k_t^2,\mu^2) = exp\left(-\int_{k_t^2}^{\mu^2}
\frac{\mathrm{d}\kappa_t^2}{\kappa_t^2}
\frac{\alpha_{s}(\kappa_t^2)}{2 \pi}\int_{0}^1
P_{qq}^{LO}(\xi)\Theta(1 - \xi -\Delta)\;\mathrm{d}\xi \right)
,
\end{equation}
\begin{equation}
\begin{split}
f_g^{\textrm{DMRW}}(x, z, k_t^2, \mu^2) = T_g^{\textrm{DMRW}}(k_t^2,
\mu^2) \frac{\alpha_{s}(k_t^2)}{2\pi}\bigg[
P_{gg}^{LO}(z)\frac{x}{z}g^{LO}(\frac{x}{z}, k_t^2)\Theta(1 - z-\Delta)
\\+ \sum_q P_{gq}^{LO}(z)\frac{x}{z} q^{LO}(\frac{x}{z},
k_t^2)\bigg],
\end{split}
\end{equation}
\begin{equation}
\begin{split}
T_g^{\textrm{DMRW}}(k_t^2,\mu^2) = exp\bigg(-\int_{k_t^{2}}^{\mu^2}
\frac{\mathrm{d}\kappa_t^2}{\kappa_t^2}
\frac{\alpha_{s}(\kappa_t^2)}{2 \pi} \int_{0}^1 \big(\xi
P_{gg}^{LO}(\xi)\Theta(1 - \xi-\Delta)\Theta(\xi-\Delta)\\ + n_f
P_{qg}^{LO}(\xi) \big)\;\mathrm{d}\xi \bigg),
\end{split}
\end{equation}
where $q={u,\overline{u}, d, \overline{d},\ldots}$.

Finally, altering the scale of the DUPDFs in the arguments of the
PDFs and the strong coupling constant in the LO-MRW approach from
$k_t^2$ to $k^2$, the modified LO-MRW DUPDFs (refer to DMRW$^\prime$ in our results) is derived and
allows us to write the following formulas for the (anti) quark and
the gluon distributions:
\begin{equation}\label{eq:thirty_one}
\begin{split}
    f_q^{\textrm{DMRW$^\prime$ }}(x, z, k_t^2,\mu^2)= T_q^{\textrm{DMRW$^\prime$}}(k^2,\mu^2)\;\frac{{\alpha_{s}(k^2)}}{2\pi}
    \bigg[ P_{qq}^{LO}(z)\frac{x}{z}q^{LO}(\frac{x}{z}, k^2)\Theta(1- z-\Delta)  \\  + P_{qg}^{LO}(z)
    \frac{x}{z} g^{LO}(\frac{x}{z}, k^2)\bigg]\;\Theta(1-z-k_t^2/\mu^2),
\end{split}
\end{equation}
\begin{equation}\label{eq:thirty_two}
T_q^{\textrm{DMRW$^\prime$ }}(k^2,\mu^2) = exp\left(-\int_{k^2}^{\mu^2}
\frac{\mathrm{d}\kappa^2}{\kappa^2} \frac{\alpha_{s}(\kappa^2)}{2
\pi}\int_{0}^1 P_{qq}^{LO}(\xi)\Theta(1 - \xi
-\Delta)\;\mathrm{d}\xi \right) ,
\end{equation}
\begin{equation}\label{eq:thirty_three}
\begin{split}
    f_g^{\textrm{DMRW$^\prime$ }}(x, z, k_t^2,\mu^2)= T_g^{\textrm{DMRW$^\prime$ }}(k^2,\mu^2)\;\frac{{\alpha_{s}(k^2)}}{2\pi}\bigg[ P_{gg}^{LO}(z)
\frac{x}{z}g^{LO}(\frac{x}{z}, k^2)\Theta(1- z-\Delta) \\ + \sum_q
P_{gq}^{LO}(z) \frac{x}{z} q^{LO}(\frac{x}{z},
k^2)\bigg]\;\Theta(1-z-k_t^2/\mu^2),
\end{split}
\end{equation}
\begin{equation}\label{eq:thirty_four}
\begin{split}
T_g^{\textrm{DMRW$^\prime$ }}(k^2,\mu^2) = exp\bigg(-\int_{k^{2}}^{\mu^2}
\frac{\mathrm{d}\kappa^2}{\kappa^2} \frac{\alpha_{s}(\kappa^2)}{2
\pi} \int_{0}^1 \big(\xi P_{gg}^{LO}(\xi)\Theta(1 -
\xi-\Delta)\Theta(\xi-\Delta)\\ + n_f P_{qg}^{LO}(\xi)
\big)\;\mathrm{d}\xi \bigg),
\end{split}
\end{equation}
respectively, where $q={u,\overline{u}, d, \overline{d},\ldots}$,
and $\Delta$ is defined as the \cref{eq:five}. In the above
equations $\kappa^2={\kappa_t^2\over {1-\xi}}$.

The resulting equations for the modified approach (DMRW$^\prime$) are the same as
the equations ($27)-(30$), with $k_t$ replaced by $k$ and also
adding a $\Theta(\mu^2-k^2)$ to prevent $k^2$ becomes larger than
$\mu^2$. In contrast to the NLO-MRW UPDFs, it is observed in the
  DMRW$^\prime$  formulas
that the LO splitting functions and the LO input PDFs are used.
The reason for such choices is that we intend to check the effects
of choosing the virtuality $k^2$ in the arguments of the PDFs and
the strong coupling constant and make a comparison between this
approach and the KMR and MRW DUPDFs  (DKMR and DMRW) approches.
\subsection{The $(z,k_t)$-factorization approach in the inclusive jet and dijet differential cross sections}
\begin{figure}
\includegraphics[width=4cm, height=4cm]{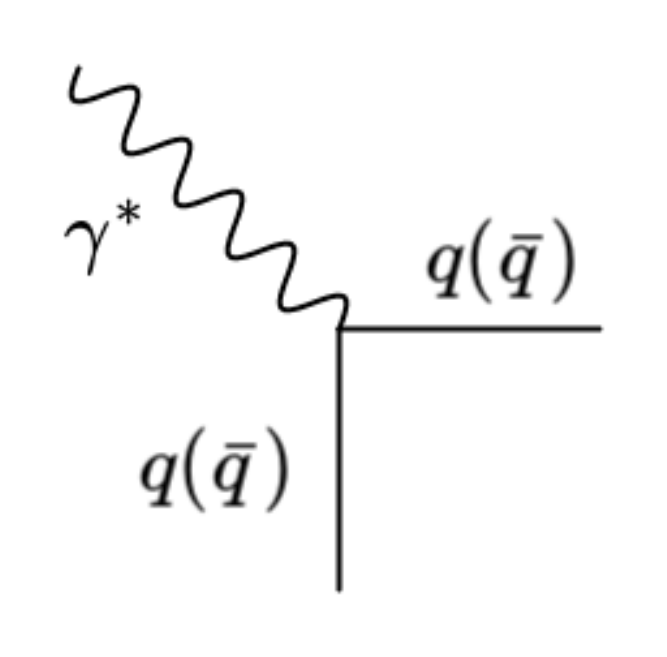}
\includegraphics[width=4cm, height=4cm]{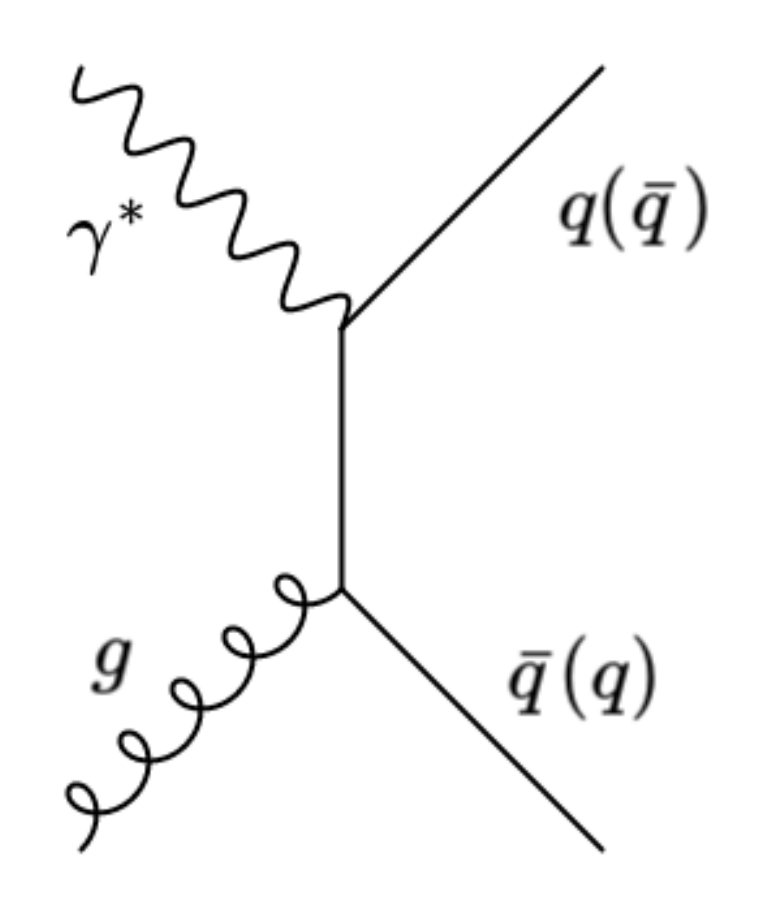}
\includegraphics[width=4cm, height=4cm]{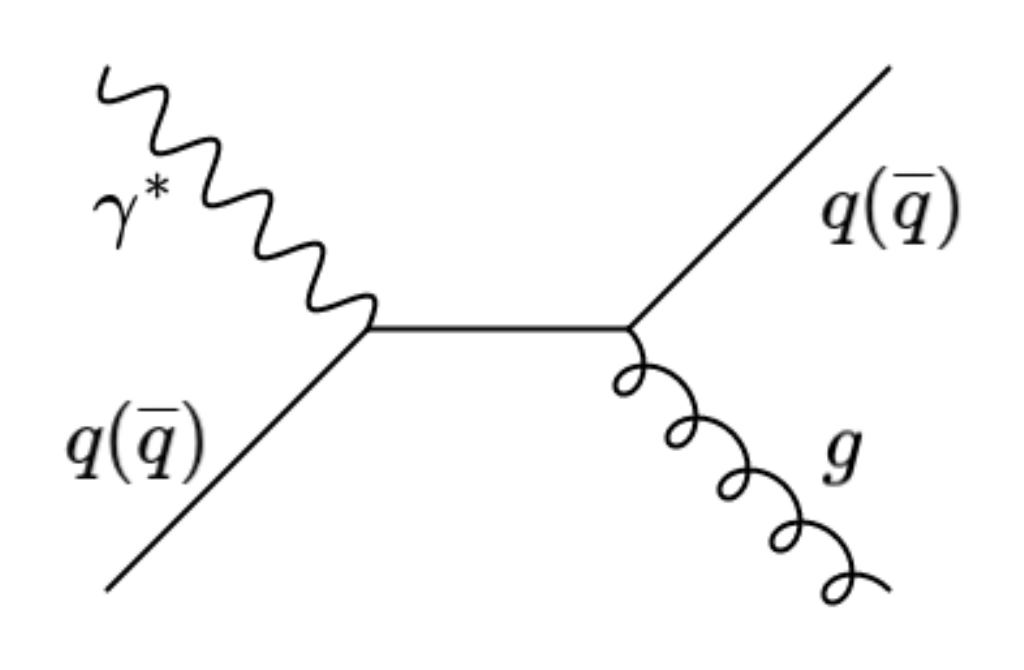}
\includegraphics[width=4cm, height=4cm]{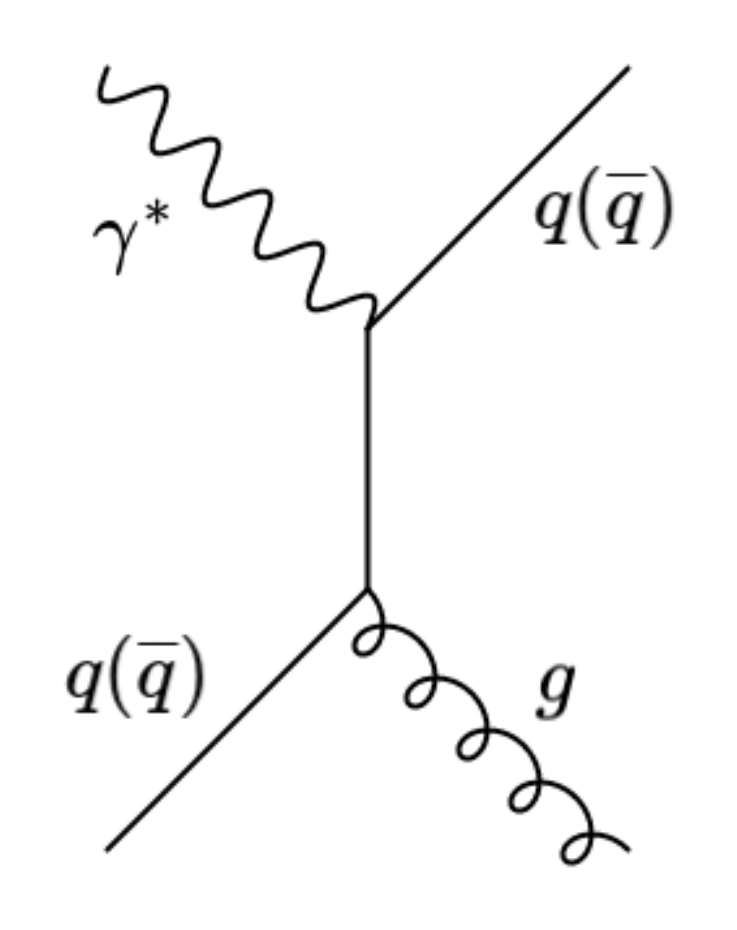}
\caption
{The  panels from left to right  shows the  sub-processes, $\gamma^\ast + q \to q$ (Born level), 
$\gamma^\ast + g \to q + \bar{q}$ (boson-gluon), $\gamma^\ast + q \to q + g$ (Compton) and $\gamma^\ast + q \to q + g$ (Compton).}
\label{fig:1}
\end{figure}

In this part of the paper, we are going to give a brief review of
some important details about the partonic level cross section,
$\sigma^{\gamma^{\ast} q} $ subprocess, see the figure \ref{fig:1}, in the
$(z,k_t)\textrm{-factorization}$ framework  \cite{wattDoubleJet}.

In the figure \ref{fig:2}, the $(z,k_t)$-factorization prescription is schematically
shown with respect to its the corresponding partonic cross section $\hat{\sigma}(x, z, k_t^2, \mu^2)$. In this framework a parton with momentum $k_{n-1}$ that
evolves according
    to the DGLAP evolution equation, with the factorization  scale $k_t$, is
    emitted a parton with momentum $k^\prime_n=k_{n-1}-k_{n}$ and becomes $k_n$, and enters into the hard
    interaction with the virtual photon, which has virtuality $Q^2$, where $q^2=-Q^2$. It can be shown that \cite{wattDoubleJet},  these diagrams with quark (gluon) that comes
    into hard interaction (as can be seen in the upper panel (lower panel) of the figure \ref{fig:2}),
    can actually in the leading logarithmic approximation be factorized into the  
    DUPDFs and the hard scattering process that now
    depends to additional fractional momentum $z$, i.e., $\hat{\sigma}^{\gamma^*{q(g)}}(x, z, k_t^2, \mu^2)$.
    In this work, we only consider the Born level sub-process, see the left panel of the figure \ref{fig:1},
     where the incoming parton has the momentum $k_n$, which is actually a dijet process in the leading
      logarithm approximation as  we said before. One of these jets, i.e., current jet, has the momentum
      $k_n + q$ and the other one which is emitted in the last step has the momentum $k_n^\prime$. Using the
      Sudakov decomposition, the four-vector $k_n$ can be described in terms of two light-like
      vectors $P$, the momentum of  proton, and $q^\prime \equiv q + x_{B} P$, in addition to the
       space-like transverse momentum, $k_T$, where the following relations are hold:
        \begin{figure} 
\includegraphics[width=15cm, height=5cm]{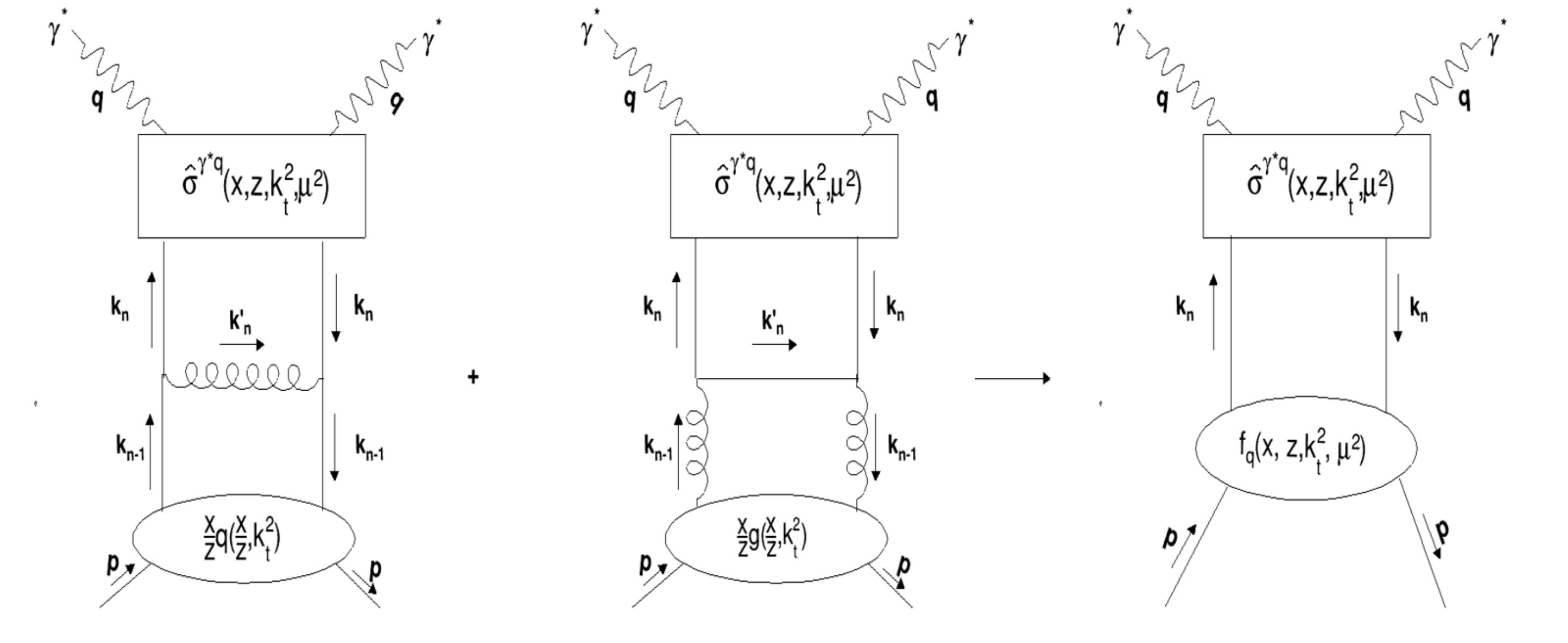}
\includegraphics[width=15cm, height=5cm]{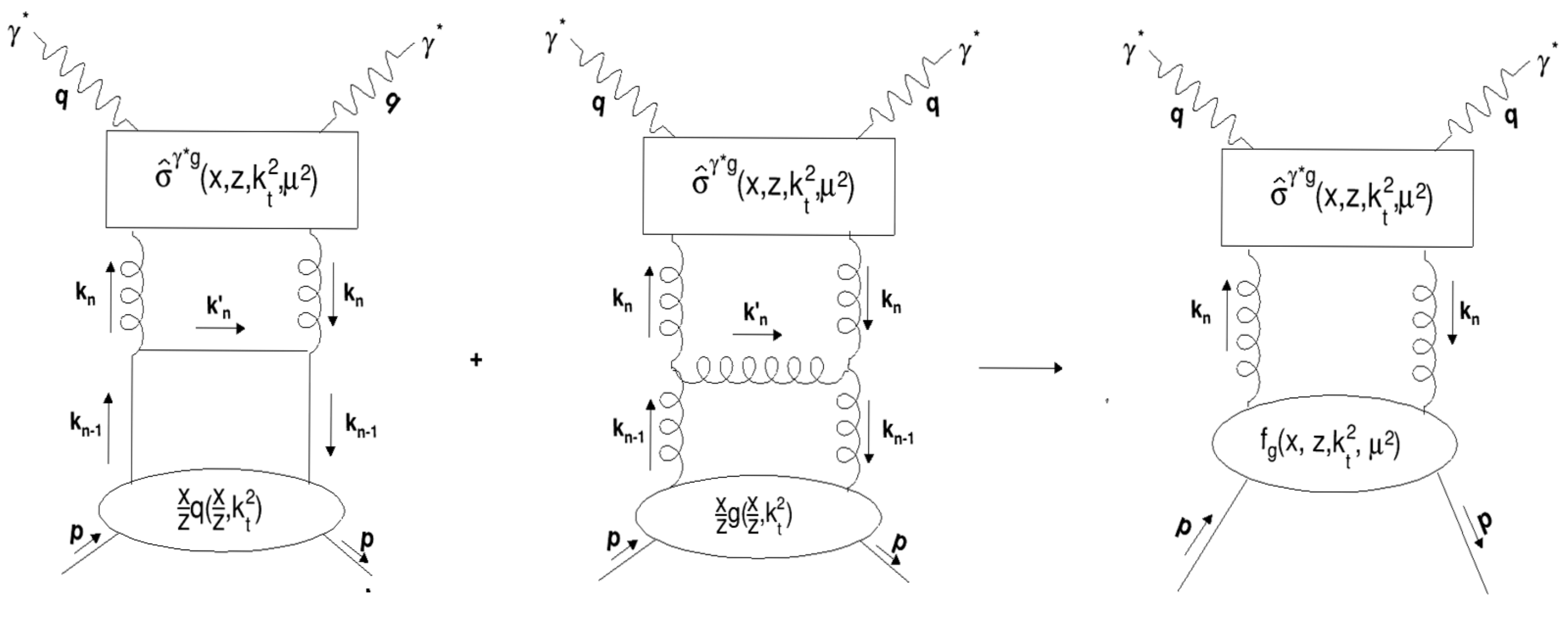}
\caption{The pictorial representation of the $(z,k_t)$-factorization for the  quark DUPDF (the three diagrams in the upper panel) and the gluon DUPDF (the three diagrams in the lower panel). The first two diagrams in the upper and lower panels show a collinear parton with four-momentum $k_{n-1}$ that emits a parton and its four-momentum becomes $k_n$ \cite{wattDoubleJet}.}
\label{fig:2}
\end{figure}     
\begin{equation}
\label{Sudakov-relations1} P^2=q^{{\prime}2}=P.k_T=q^\prime.k_T=0
\end{equation}
\begin{equation}
\label{Sudakov-relations2} k_T^2=-k_t^2,\;\;\;\;\; P.q^\prime =
\dfrac{Q^2}{2 x_{Bj}}
\end{equation}

Therefore the momenta of $k_n$ and $k_{n-1}$ can be written as
follows:
\begin{equation}
\label{incoming-parton-momentum} k_n = xP - \beta q^{\prime} +
k_{T},
\end{equation}
\begin{equation}
k_{n-1} = \dfrac{x}{z} P.
\end{equation}
One can obtain $\beta$ from the \cref{incoming-parton-momentum},
using the on-shell condition of the emitted parton in the last
step of evolution chain, i.e.,:
\begin{equation}
{k_n^\prime}^2 = (k_{n-1} - k_n)^2 = 0 \implies \beta = \dfrac{z k_t^2
x_{Bj}}{x(1-z)Q^2}.
\end{equation}
Here, the full kinematic of the parton in the last step of the
evolution is taken into account, and one can see that $k_n^2 =
k_t^2/(1-z)$, in contrast to the $k_t$-factorization framework,
where $z\to 0$ and as a result of this, one obtains $k = x P +
k_t$, and $k^2 = -k_t^2$. Additionally, using the constraint
$(k_n+q)^2=0$, one can obtain the following relation for $x$:
\begin{equation}
\label{frac_momenta_zkt}
x_{\pm}=\dfrac{x_{Bj}}{2(1-z)}\left(1-z+\frac{k_t^2}{Q^2}
\pm\sqrt{{(1-z+\frac{k_t^2}{Q^2})}^2-4\frac{k_t^2}{Q^2}z(1-z)}
\right).
\end{equation}
In order to calculate the   cross section one need to use the momenta of
partons in the Briet frame. In this frame, the virtual photon only has
component of momentum along the $z$ direction $q=(0,0,0,-Q)$. Then if we  use the above
 \cref{Sudakov-relations1,Sudakov-relations2}, and utilize the light-cone components, the momenta can be written as follows:
\begin{equation}
p = \dfrac{1}{\sqrt{2}}(Q/x_{Bj},0,\boldsymbol{0}), \;\;\;\;\;
q^\prime =\dfrac{1}{\sqrt{2}} (0, Q, \boldsymbol{0}), \;\;\;\;\;
k_T=(0,0,\boldsymbol{k_t}),
\end{equation}
where we use the definition of $a=(a^+,a^-,a_T)$, with
$a^+=(a_0+a_3)/\sqrt{2}$ and $a^-=(a_0-a_3)/\sqrt{2}$ for the light-cone variables \cite{Collins:1997nm}. Now, these momenta can be
used to obtain the following rapidities for the current jet:
\begin{equation}
\eta^{Breit} =\dfrac{1}{2}\ln{\dfrac{k_n^+ + q^+}{k_n^-+q^-}} =
\dfrac{1}{2} \log{\dfrac{x/x_B-1}{1- \beta}},
\end{equation}
and the last step emitted jet:
\begin{equation}
\eta^{Breit} =\dfrac{1}{2}\ln{\dfrac{k_{n-1}^+ - k_n^+}{k_{n-1}^-
- k_n^-}} =  \dfrac{1}{2} \log{\dfrac{x(1-z)}{x_B z \beta}}.
\end{equation}

To calculate the hadronic cross section, first, one needs to obtain 
the partonic cross section \cite{wattDoubleJet}, i.e.,
\begin{equation}
\label{parton-cs-zkt} d \hat{\sigma}^{\gamma^{\ast} q^\ast}_{T, L}
= d\Phi^{\gamma^\ast q^\ast} |M^{\gamma^\ast q^\ast}_{T,
L}|^2/F^{\gamma^\ast q^\ast},
\end{equation}
where $T$ and $L$ are denoted for the longitudinal and transverse
virtual photon polarizations cross section. The flux factor and the
phase space can be obtained as:
\begin{equation}
F^{\gamma^\ast q^\ast} = 4xp.q=2xQ^2/x_{Bj},
\end{equation}
\begin{equation}
\label{phase-space} d\Phi^{\gamma^\ast q^\ast} = 2\pi
\dfrac{x_{Bj}}{Q^2} \sum_{i=\pm}\dfrac{1}{(1-x_{Bj}
\beta/x)}\delta(x-x_i),
\end{equation}
 respectively, where $x_i$ is defined in the \cref{frac_momenta_zkt}, and because
 $x_{Bj} \le x$, it constraints $x$ to $x_+$.

The squared matrix elements in the \cref{parton-cs-zkt} are given as follows:
\begin{equation}
|M_T^{\gamma^\ast q^\ast}|^2 = 4 \pi \alpha_{em} e_q^2 Q^2
\dfrac{x}{x_{Bj}}, \;\;\;\;\;\;\;\;\;\;\;\;\;\; |M_L^{\gamma^\ast
q^\ast}|^2 = 0,
\end{equation}

Now one can insert  above different elements into \cref{parton-cs-zkt} to get
the partonic cross sections as:
\begin{equation}
\hat{\sigma}_T^{\gamma^*q^\ast}(x,z,k_t^2,\mu^2) =
\frac{4\pi^2\alpha_{em}}{Q^2}e_q^2 \dfrac{x_B}{1-x_B{\beta/x}}\delta(x-x_+)
\;\;\; with \;\;\;
\hat{\sigma}_L^{\gamma^*q^\ast}(x,z,k_t^2,\mu^2) =0.
\end{equation}
After using the above cross section in the \cref{eq:twenty_two},
the hadronic cross section can be obtained, i.e,:
\begin{equation}
\label{eq:thirty_eight} \sigma_T^{\gamma^*p}
=\frac{4\pi^2\alpha}{Q^2}
\int_x^1\;dz\;\int_0^\infty\frac{\mathrm{d}k_t^2}{k_t^2}\dfrac{{x_B}/x}{1-x_B{\beta/x}}\sum_q\;e_q^2f_q(x,z,k_t^2,\mu^2).
\end{equation}
Finally, using the above equation, the calculation of inclusive
jet and dijet production in the electron-proton collision can be
evaluated as:
\begin{equation}
\begin{split}
\sigma(jet) =
\int_{y_{min}}^{y_{max}}\;dy\;\int_{Q^2_{min}}^{Q^2_{max}}\;dQ^2\dfrac{\alpha}{2\pi
y Q^2}[(1+(1-y)^2 \sigma_T^{\gamma^*p}+2(1-y)\sigma_L^{\gamma^*p}]
\\
\sum_{jets}\;\Theta(E_T-E_{T_{min}})\;\Theta(E_{T_{max}}-E_T)\;\Theta(\eta-\eta_{min})\;\Theta(\eta_{max}-\eta),
\end{split}
\end{equation}
where $y$, $Q^2$, $E_T$, $\eta$ are the inelasticity, the
virtuality, the transverse energy of   produced jets, and the
pseudo-rapidity, respectively. Then we can  calculate the above differential
cross section  \cite{wattDoubleJet},
  as:
\begin{equation}
d\sigma/dO = \sigma(jet) /(O_{max} - O_{min}),
\end{equation}
where $O$ can be any physical observable such as $E_T$, $M_{jj}$
etc.. Also the denominator in the above formula   is the size of
each bin.
\subsection{The numerical methods for the differential cross sections calculation of high $Q^2$ in the
inclusive jet and dijet} 
In contrast to the calculation of $k_t$-factorization framework of
the section II, which was done by the KaTie event generator, we
directly calculate the high $Q^2$ inclusive jet and dijet cross
sections to be compared with those of ZEUS collaboration  \cite{Zeus_2010,zeus2002}. For the numerical calculation
of these  differential cross sections, we need to generate the DUPDFs for the
three aforementioned models, see the sub-section \ref{DUPDF}, and we should also use a method
to numerically compute the multidimensional integral over $z$,
$k_t^2$, $y$ and $Q^2$. The DUPDFs integrals   are again calculated with the help of  
Gauss-Legendre quadrature method of the BOOST libraries. While  the input PDFs for the DUPDFs, the
central PDFs of MMHT2014lo68cl LHAPDF library are used. To
perform the multidimensional integrals of the differential cross
sections we resort to the VEGAS Monte-Carlo
algorithm~\cite{NUMERICAL_RECIPES}, with the limit of integrals
over $y$ and $Q^2$ and also the cuts over $\eta$ and $E_T$ are set
in accordance with the ZEUS collaboration
data~\cite{zeus2002,Zeus_2010}. The factorization scale $\mu$ is
set as \cite{DUPDFElectroweak}:
$$
    \mu = Q\dfrac{x}{x_B}\sqrt{\frac{1-\beta}{x/x_B-1}}.
$$

As we mentioned in the
section \ref{KATIE_N}, we again only include the first five quark
flavors and their anti-quarks in our calculation. Here an
important point about the Born level process, in which the jets are widely
separated from each other, is that there is a one to one relation
between the momentum of partons and jets, and hence using the jet
algorithm  for
calculation of the inclusive jet and the dijet at this level of
perturbation theory is unnecessary  \cite{wattDoubleJet}.
\section{Numerical results and discussions}\label{Numerical-Results}
In this section, our predictions of the different differential
cross sections for the inclusive jet and dijet cross sections, using the $k_t$ and
$(z,k_t)\textrm{-factorizations}$ are compared to the experimental
data of    ZEUS  collaboration  
\cite{zeus2002,Zeus_2010}.  In the inclusive jet experiment
\cite{zeus2002}, positrons of energy $27.5\;GeV$ are collided with
the protons of energy $820\;GeV$. While for the inclusive dijet
experiment \cite{Zeus_2010}, the electrons or the positrons of
energy $27.5\;GeV$ are collided with the protons of energy
$920\;GeV$. The inelasticity in the inclusive dijet experiment is
between $0.2$ and $0.6$, while in the inclusive jet experiment it
is fixed by $ -0.7 < \cos{\gamma} < 0.5$, where $\cos{\gamma}$ is
defined as:
\begin{equation}
\cos\gamma = \dfrac{x_B(1-y)E_p - y E_e}{x_B(1-y)E_p + y E_e}.
\end{equation}
The virtuality $Q^2$ for both experiments is larger than $125\;
 GeV^2$, but
    for the dijet experiment is limited to $Q^2 < 20000 \;GeV^2$. For the inclusive
     jet experiment at least one jet has the transverse energy in the Breit frame larger than
      $8\;GeV$ and $ -2<\eta^{jet}_{B}<1.8$, and for the inclusive dijet experiment the  highest transverse energies of the two
       jets     should be larger than $8\; GeV$, with $ -1 < \eta_{Lab}^{jet} < 2.5$.
       In addition to these limits, there is also one additional constraint  on the invariant mass of the final  dijet state,
       which limits the dijet invariant mass to $M_{jj} > 20\;GeV$. The tested channels of differential cross sections that we consider here are as following:
\begin{itemize}
    \item The inclusive jet cross sections:
    \begin{enumerate}
        \item $\dfrac{d\sigma}{dE_{T,B}^{jet}}$: The differential cross section with respect to the   transverse energies of the final state jets in the Breit Frame.
        \item $\dfrac{d\sigma}{d\eta_B^{jet.}}$: The differential cross section with respect to the pseudo-rapidity of the final state jets in the Breit Frame.
        \item  $\dfrac{d\sigma}{dQ^2}$: The differential cross section with respect to the virtuality.
    \end{enumerate}
        \item The inclusive dijet cross sections:
        \begin{enumerate}
            \item $\dfrac{d\sigma}{d\overline{E_{T,B}^{jet}}}$: The differential cross section with respect to the mean transverse jet energy in the Breit frame
        \item  $\dfrac{d\sigma}{dQ^2}$: The differential cross section with respect to the virtuality.
        \item  $\dfrac{d\sigma}{d\eta^{\ast}}$: The differential cross section with respect to the half difference of the jet pseudo-rapidities in the Breit frame, where $\eta^\ast$ is defined as $|\eta_{B}^{jet1} - \eta_{B}^{jet2}|/2$.
        \item $\dfrac{d\sigma}{dM_{jj}}$: The differential cross section with respect to the invariant mass of   dijet.

        \end{enumerate}
\end{itemize}

\begin{figure}
   \includegraphics[width=7cm, height=9cm]{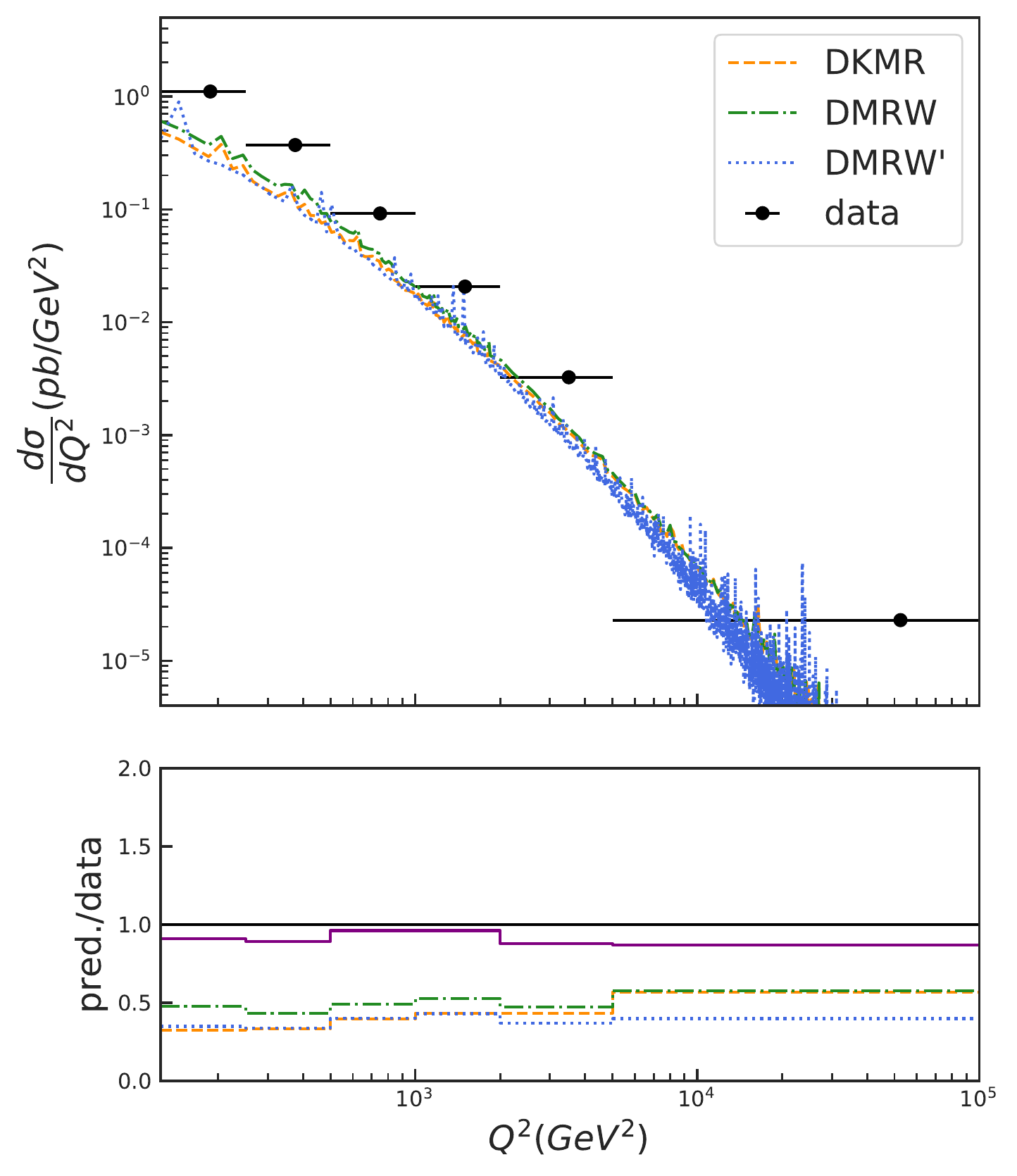}
    \includegraphics[width=7cm, height=9cm]{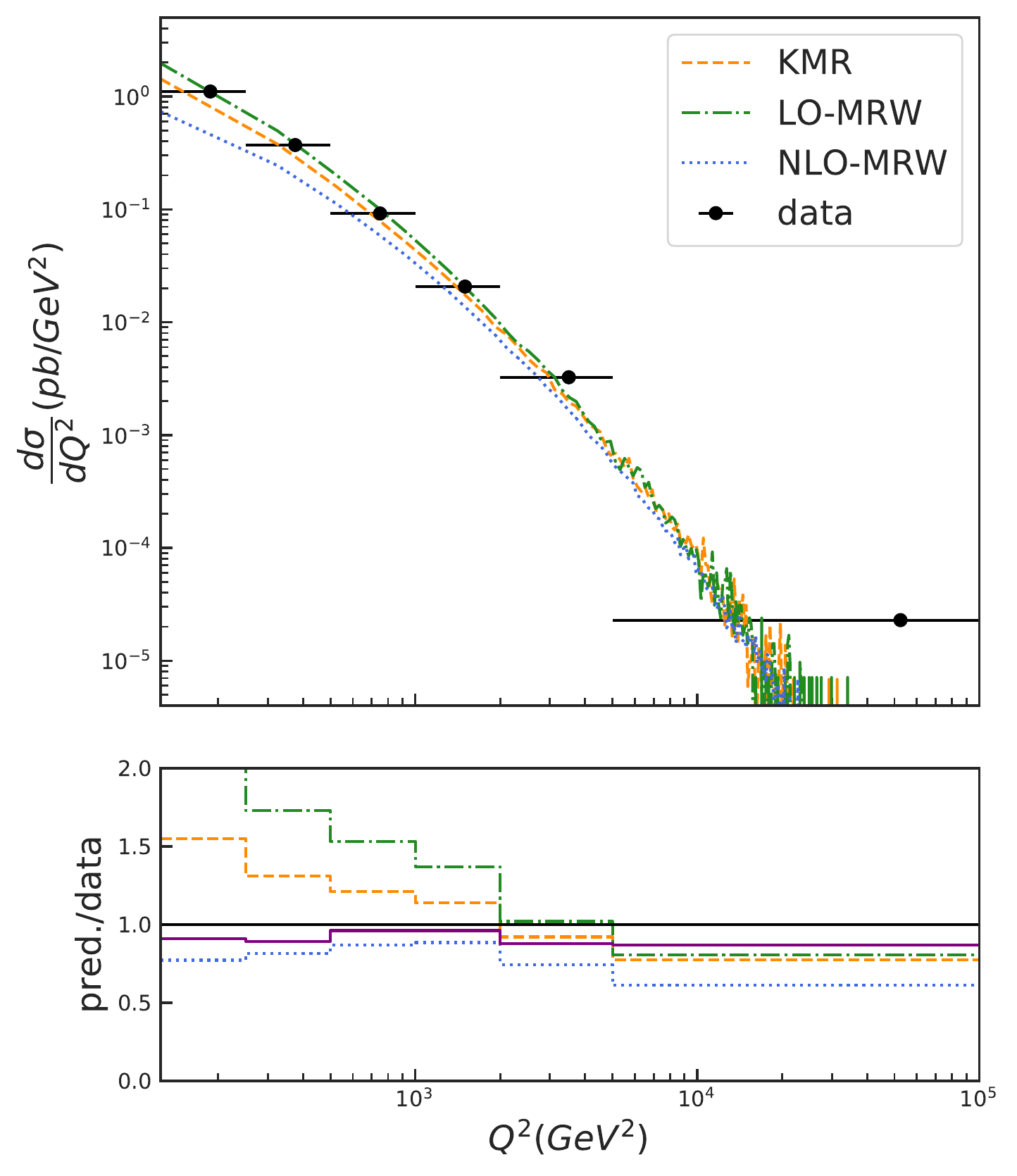}
\caption{The left and right panels show the comparison of inclusive jet production cross sections, ($d\sigma/dQ^2$), for the $(z,k_t)$-factorization with the DUPDFs models, i.e., DKMR, DMRW, and DMRW$^\prime$, and the $k_t$-factorization with the UPDFs models, i.e., KMR, LO-MRW, and the NLO-MRW with respect to the experimental data \cite{zeus2002}. The below panels demonstrate predictions ratio of our models  and  the NLO collinear (solid-purple line) to the data  \cite{zeus2002}, respectively.
        }
    \label{fig:4} 
\end{figure}

\begin{figure}
    \includegraphics[width=5cm, height=9cm]{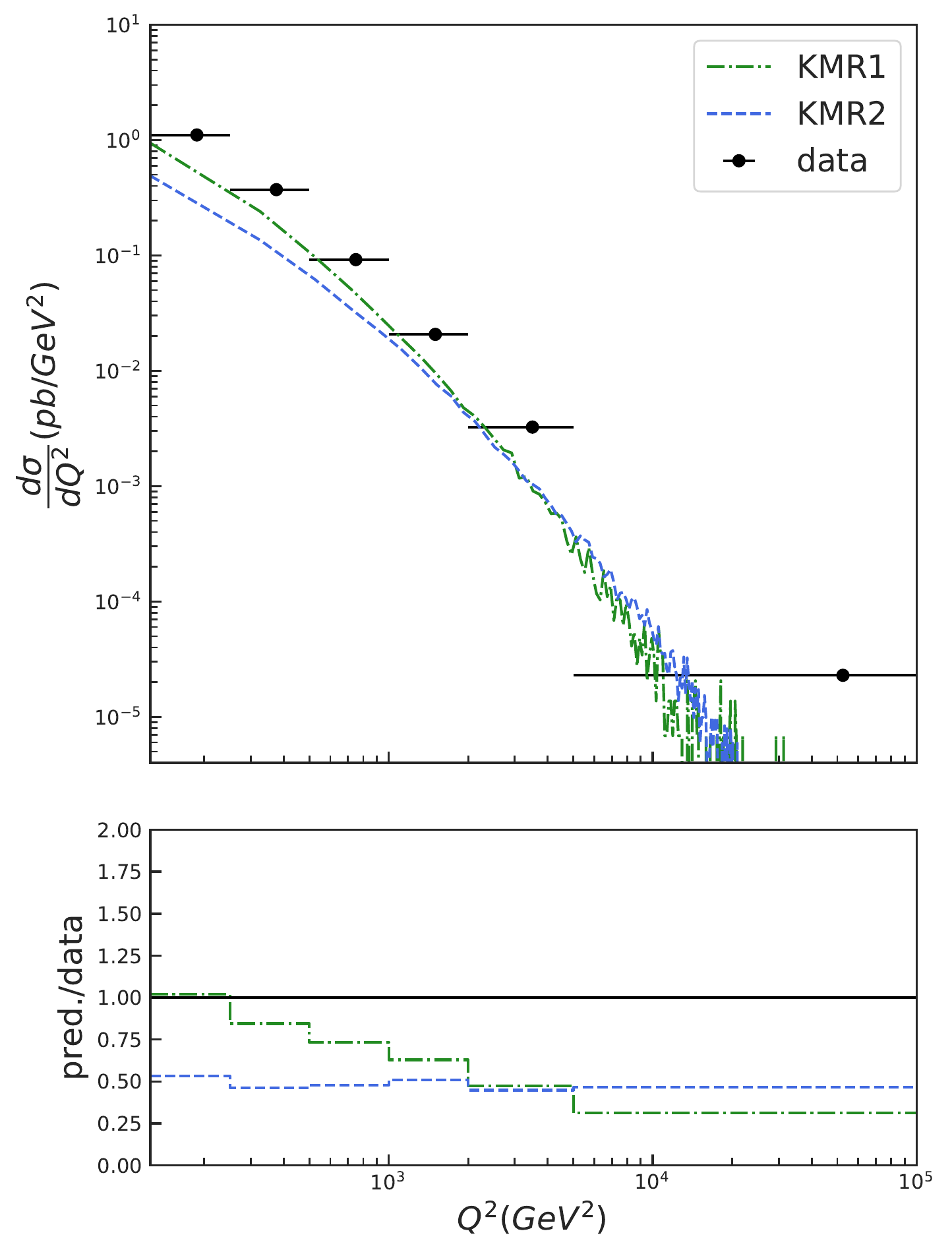}
    \includegraphics[width=5cm, height=9cm]{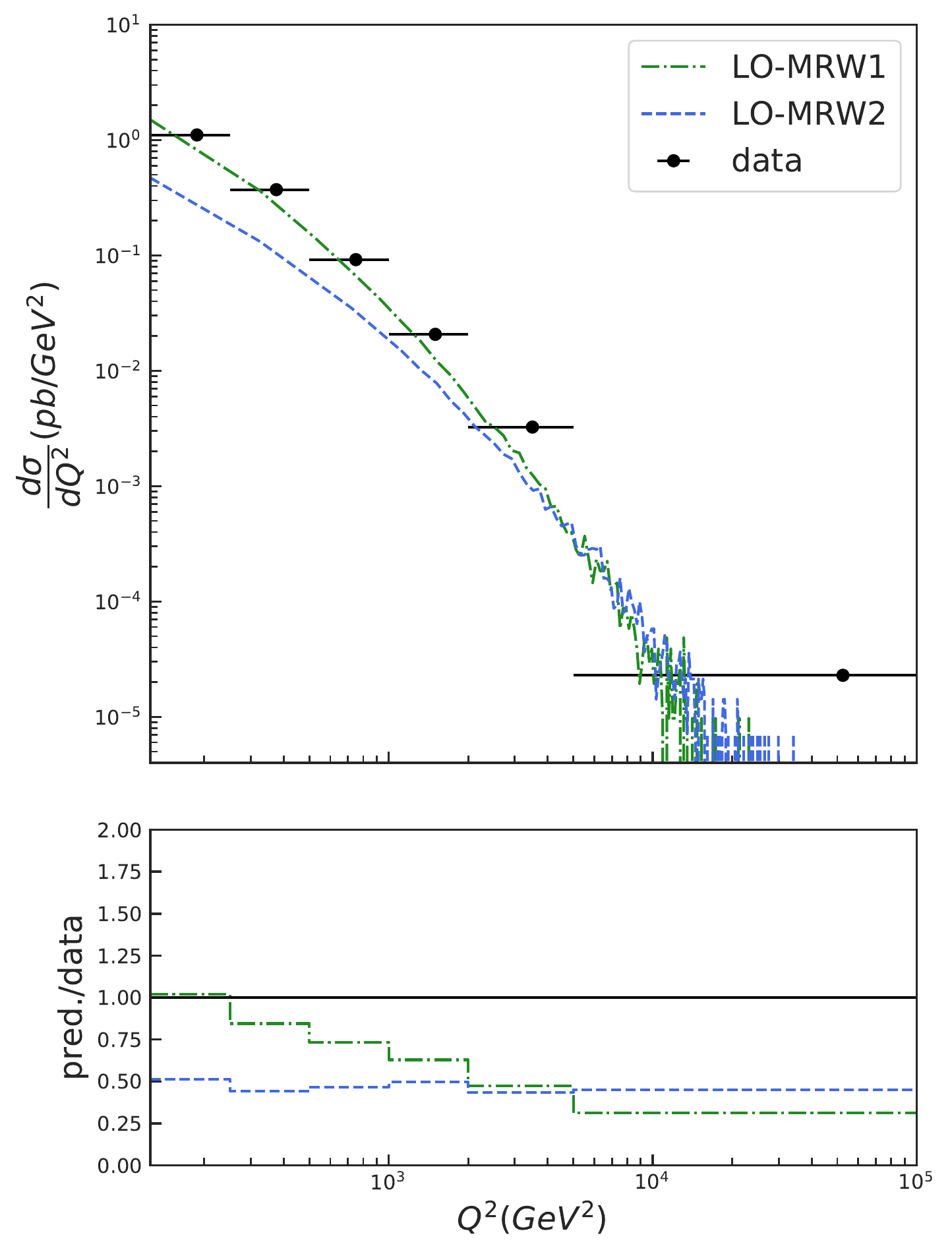}
    \includegraphics[width=5cm, height=9cm]{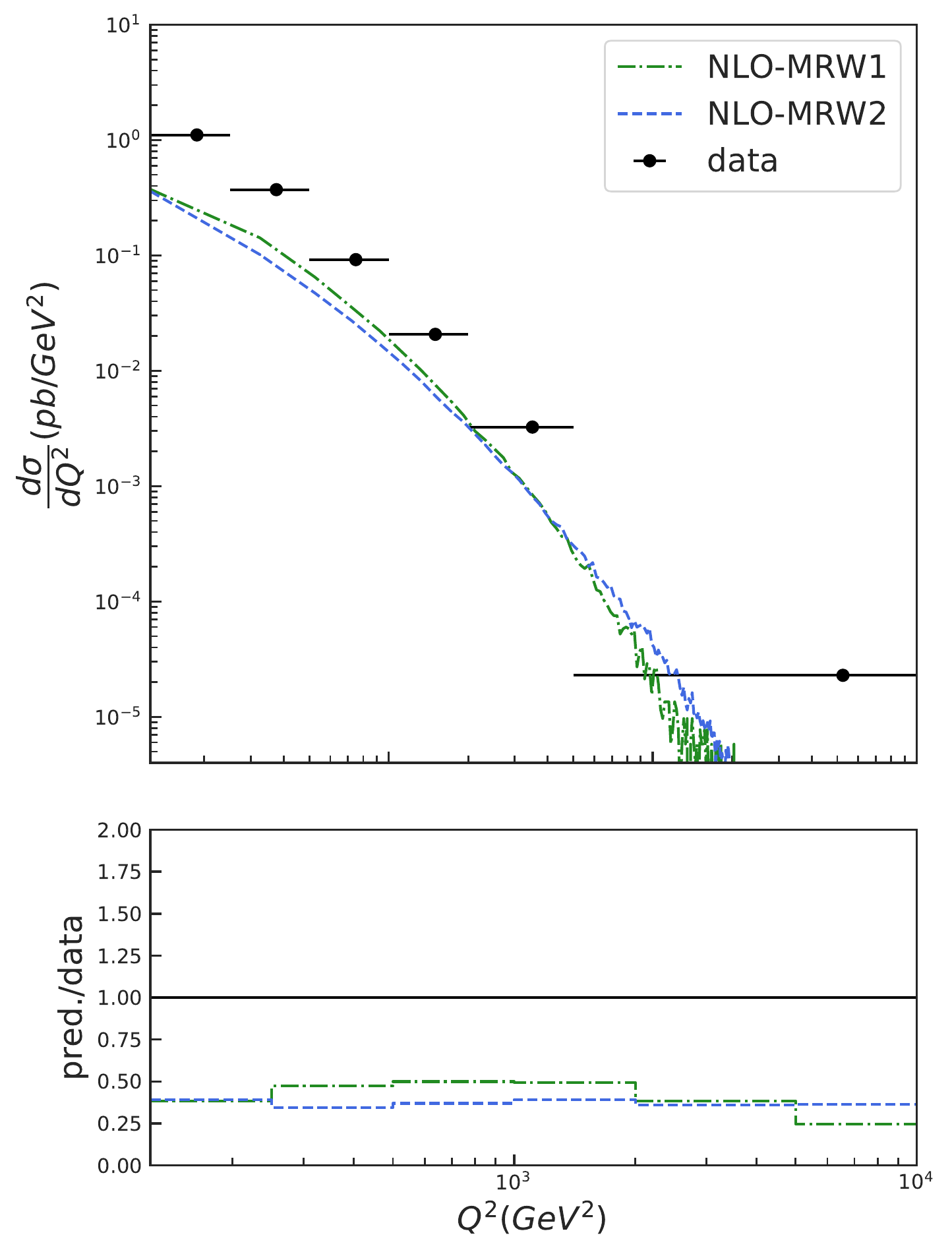}
    \caption
        {The contribution of each  sub-processes to inclusive jet differential cross section, $d \sigma/ dQ^2$,  with respect to the
            experimental data \cite{zeus2002} for the KMR, LO-MRW and the NLO-MRW, i.e.,   $\gamma^\ast + q \to q $ (Born level) (KMR1, LO-MRW1, NLO-MRW1) and         $\gamma^\ast + q \to q + g$ (boson-gluon), $\gamma^\ast + g \to q + \bar{q}$ (Compton) and $\gamma^\ast + g \to q + \bar{q}$ (Compton) (KMR2, LO-MRW2, NLO-MRW2). See the figure 1.
        }
    \label{fig:5}
\end{figure}

\begin{figure}
    \centering
    \includegraphics[width=7cm, height=9cm]{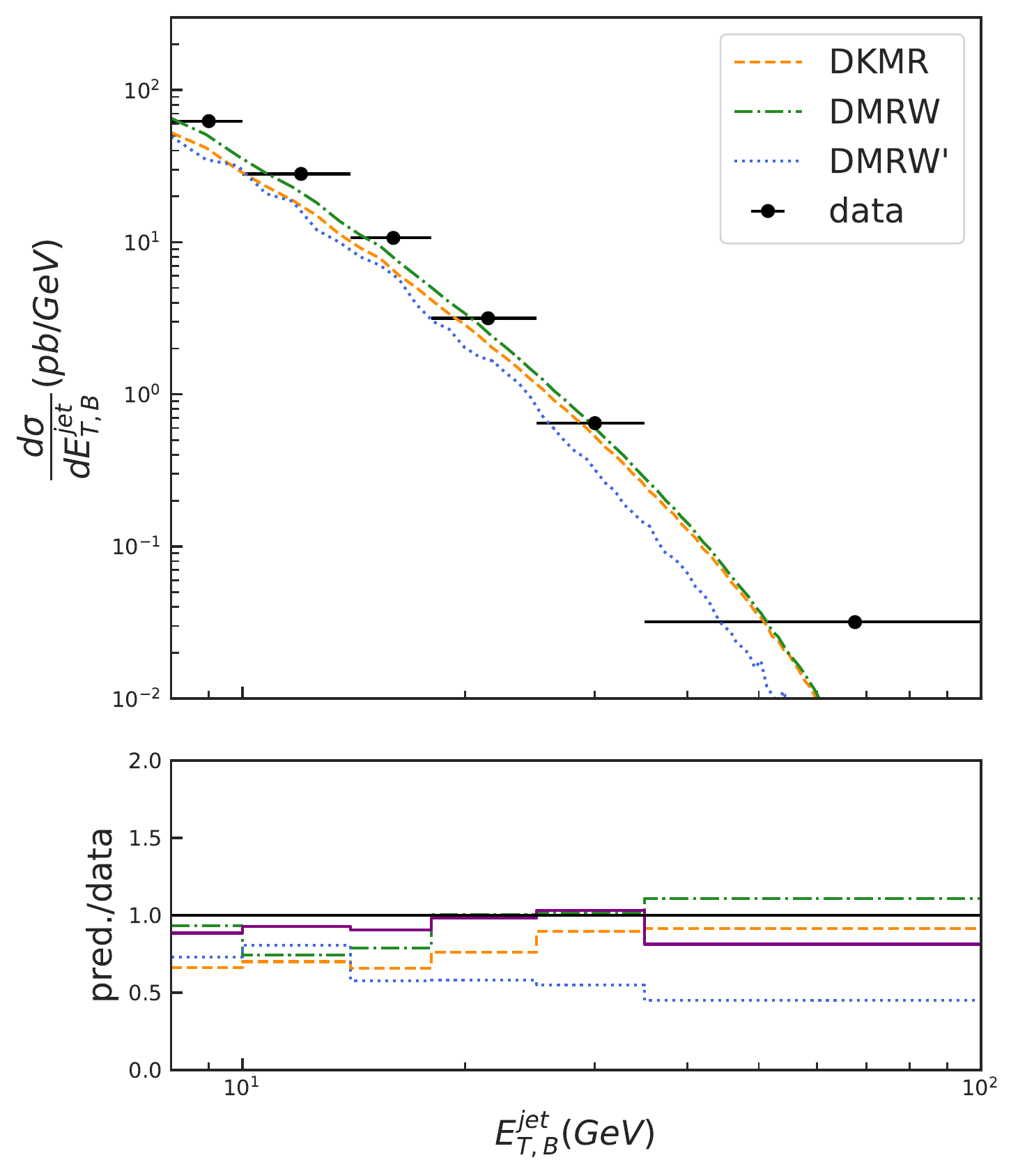}
    \includegraphics[width=7cm, height=9cm]{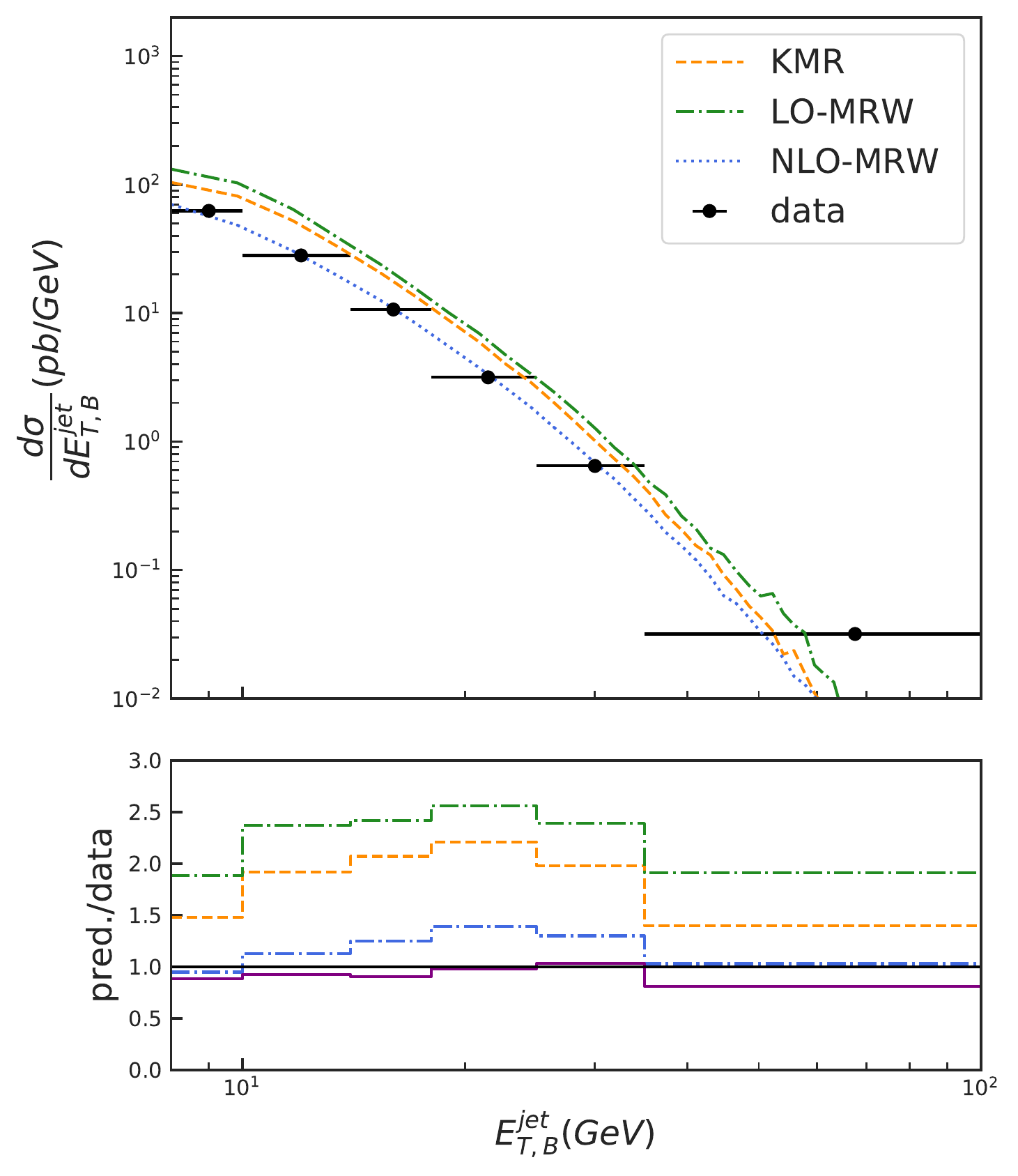}
    \caption{    The same as the figure $4$ but for $d\sigma/dE_{T,B}^{jet}$.}
     \label{fig:6}
\end{figure}

\begin{figure}
    \includegraphics[width=5cm, height=9cm]{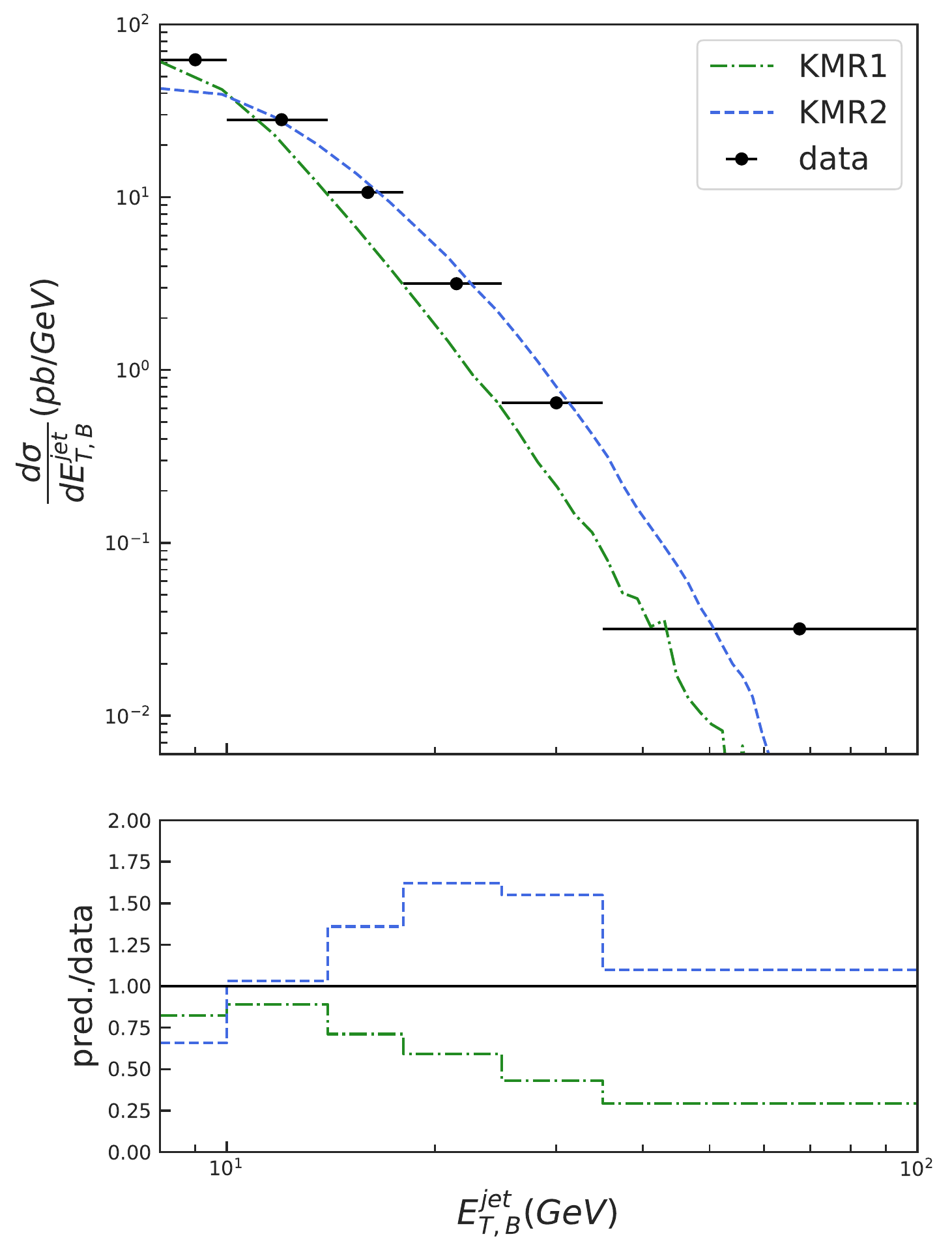}
    \includegraphics[width=5cm, height=9cm]{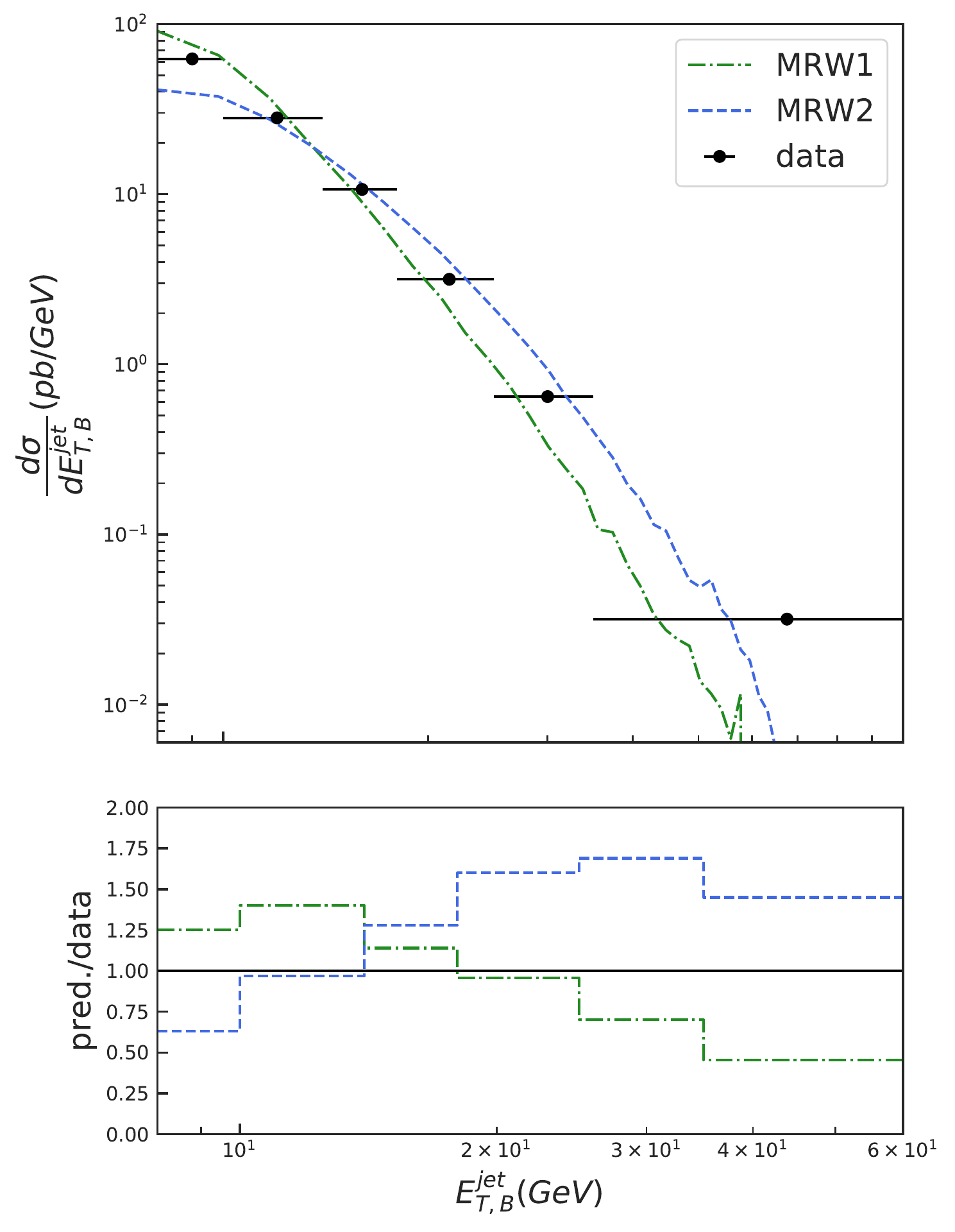}
    \includegraphics[width=5cm, height=9cm]{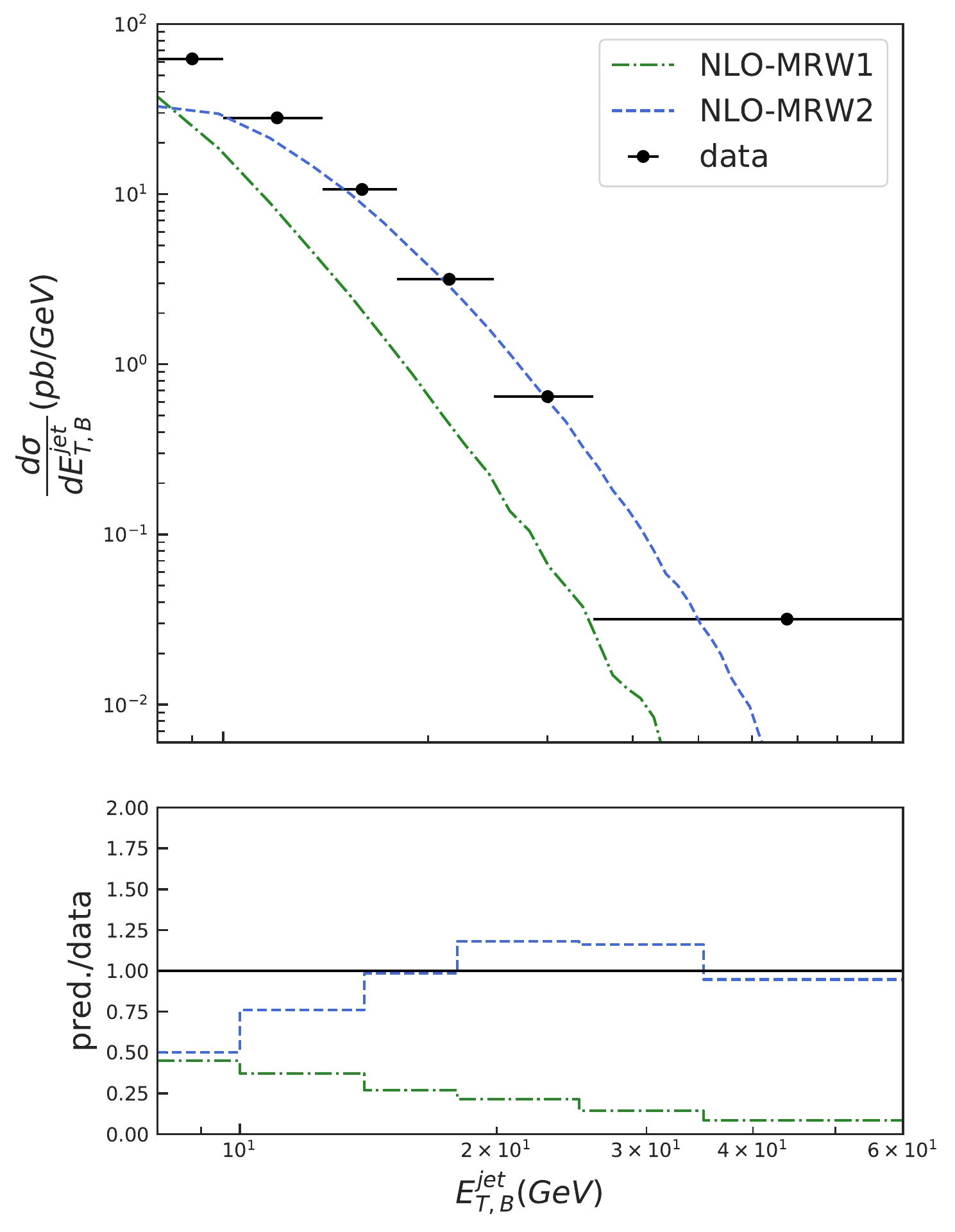}
    \caption{
            The same as the  figure $5$ but for  $d\sigma /dE_{T,B}^{jet}$.
        }
    \label{fig:7}
\end{figure}

\begin{figure}
    \includegraphics[width=7cm, height=9cm]{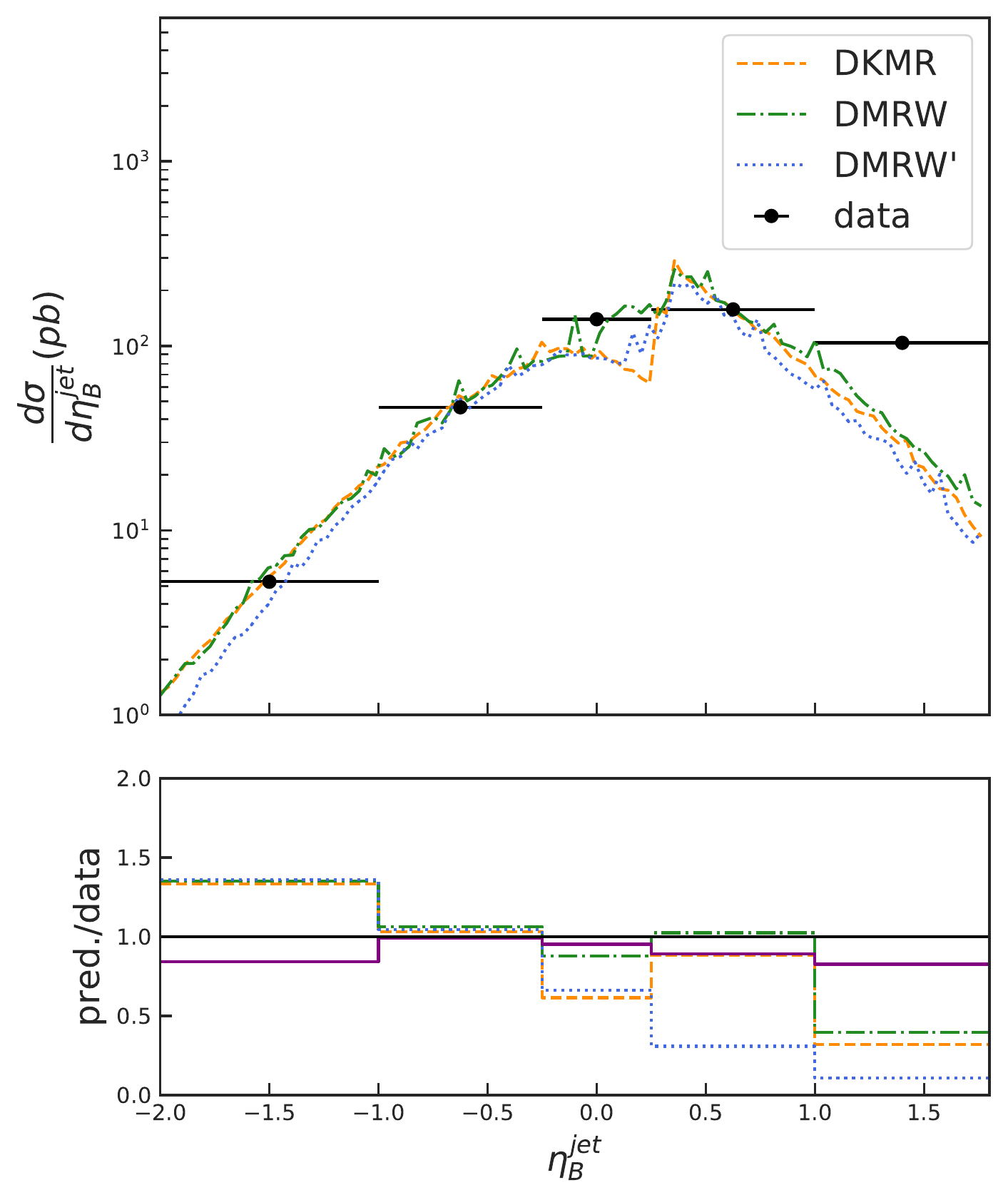}
    \includegraphics[width=7cm, height=9cm]{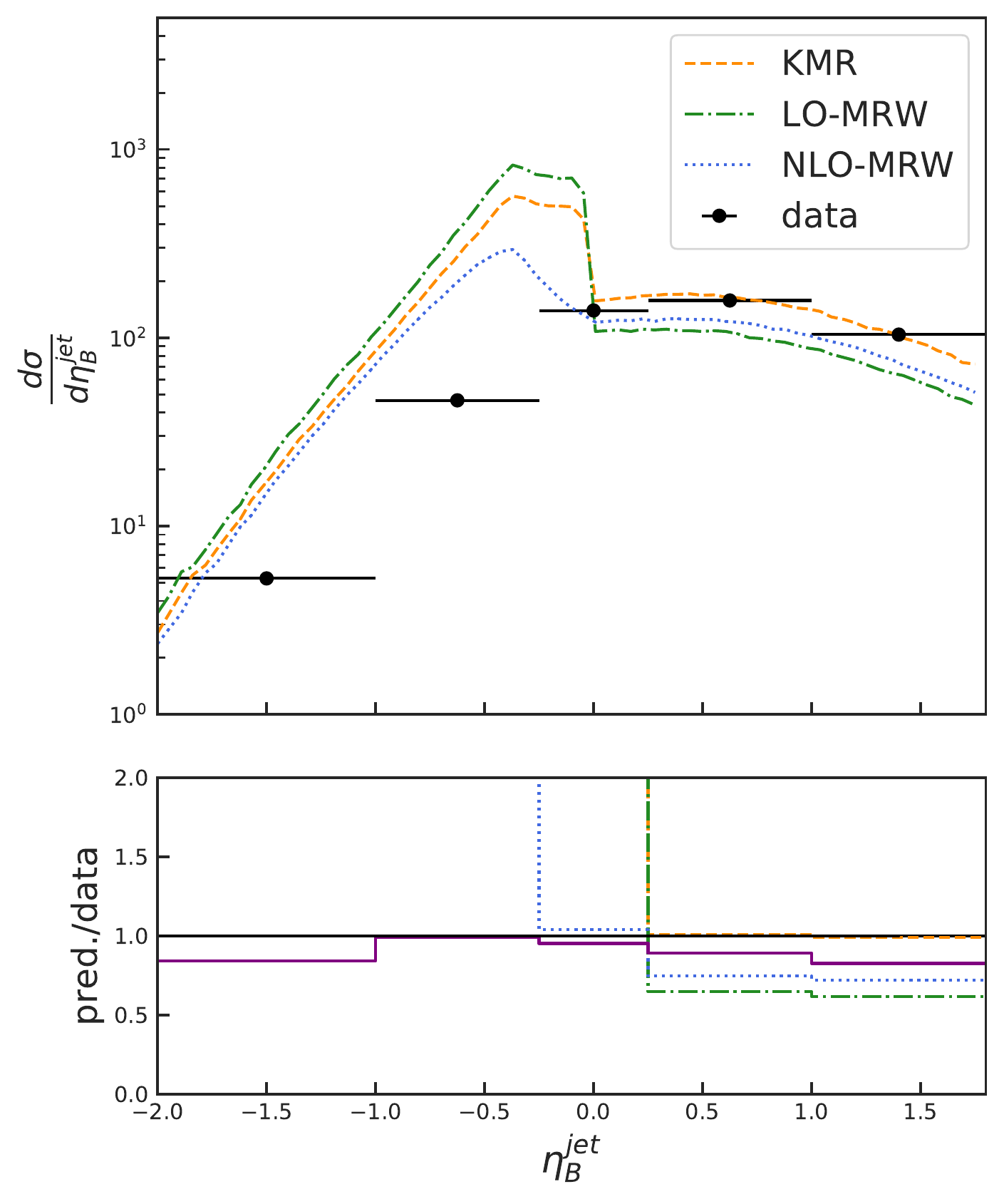}
  \caption{The same as figure $4$ but for $d\sigma /d\eta^{jet}_{Breit}$.
        }
    \label{fig:8}
\end{figure}

\begin{figure}
    \includegraphics[width=5cm, height=9cm]{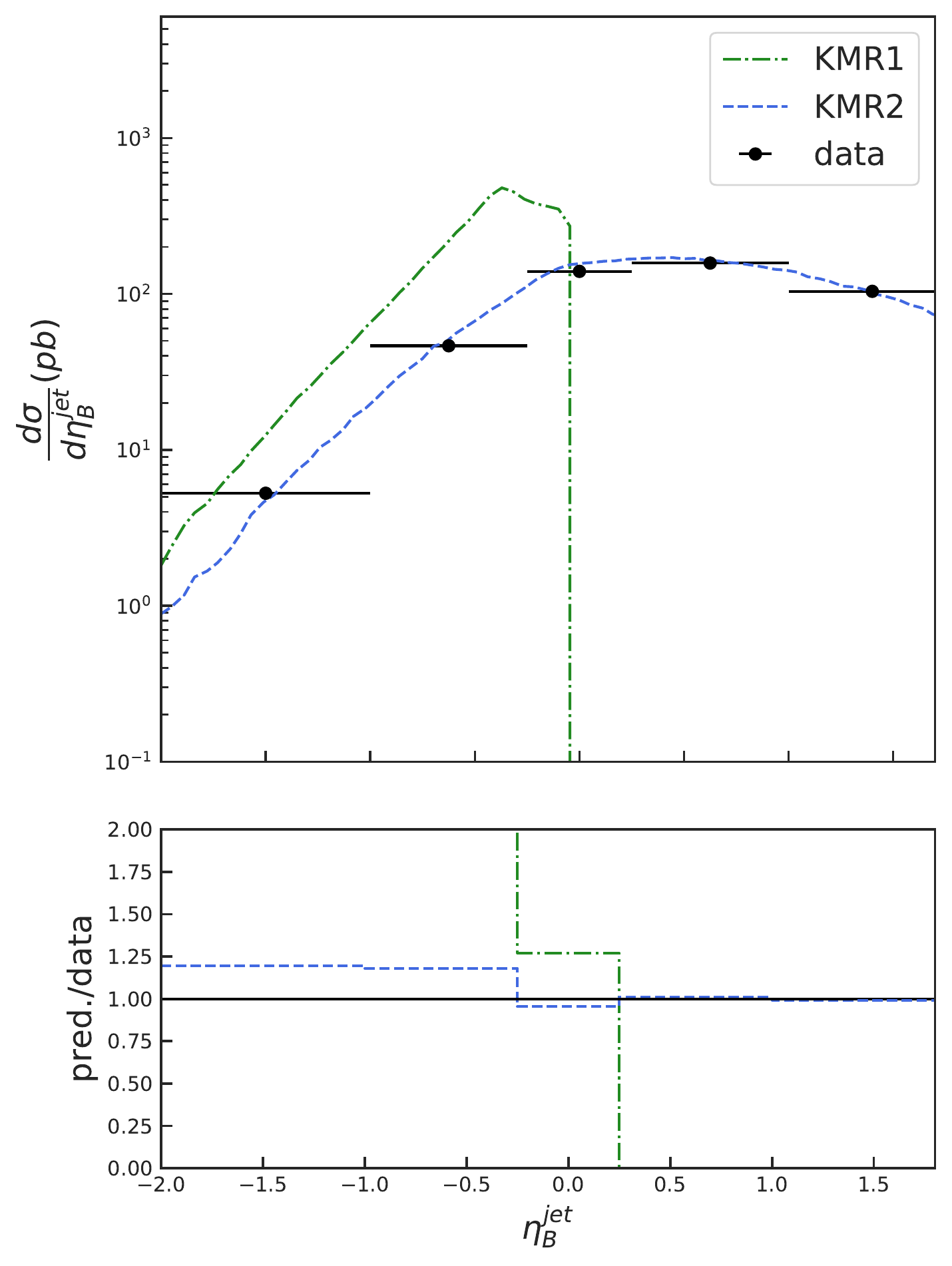}
    \includegraphics[width=5cm, height=9cm]{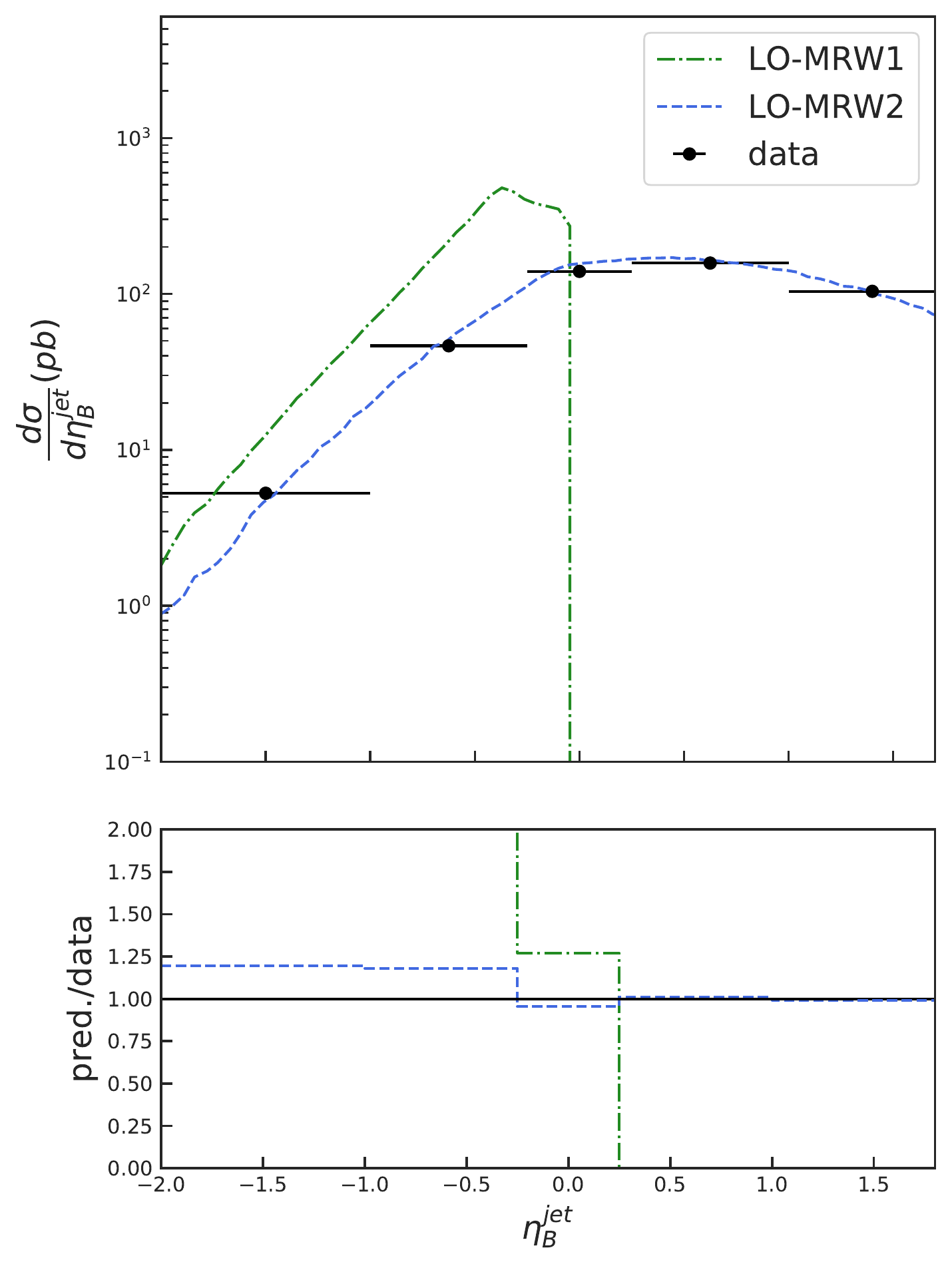}
    \includegraphics[width=5cm, height=9cm]{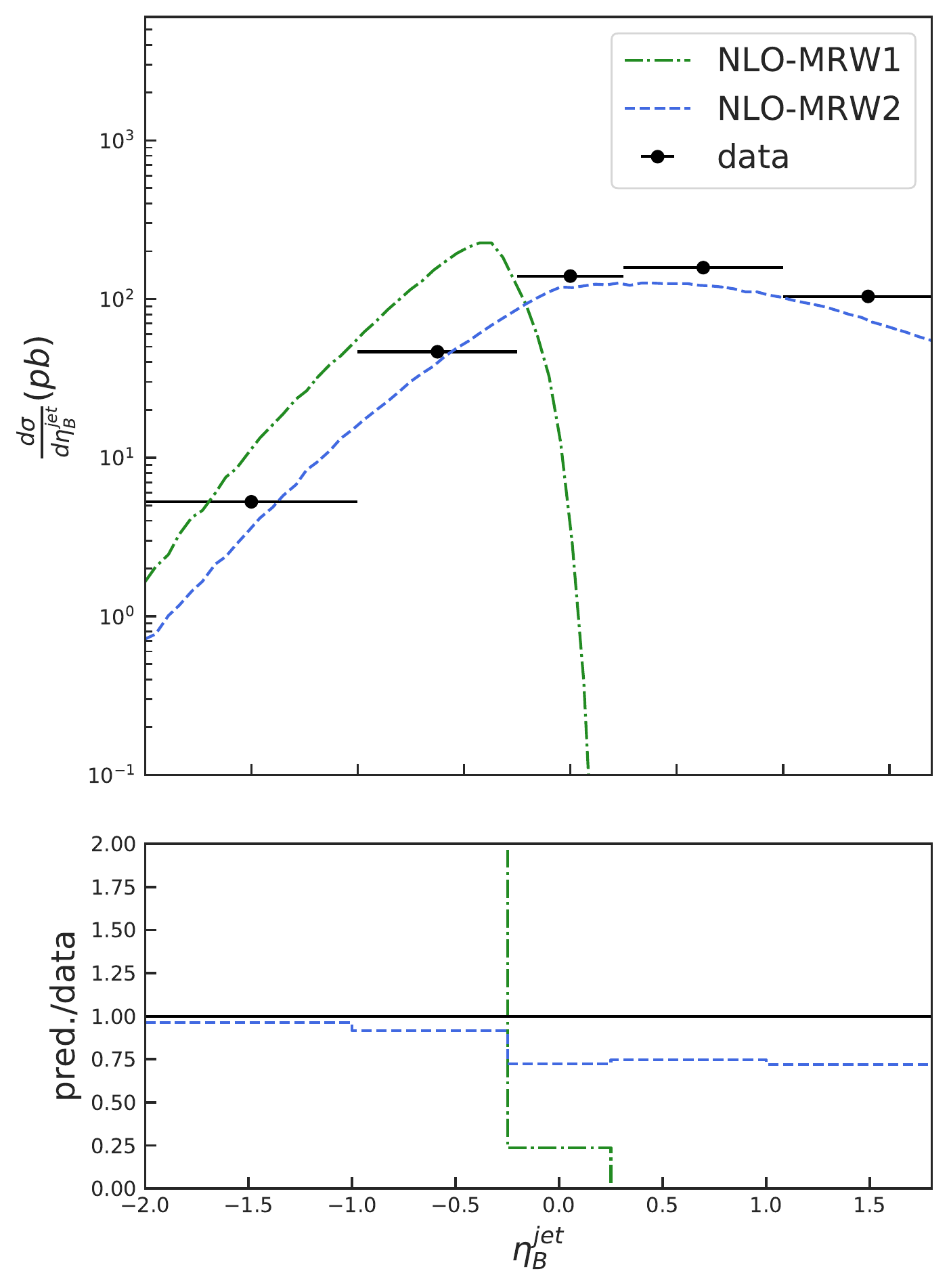}

    \includegraphics[width=5cm, height=9cm]{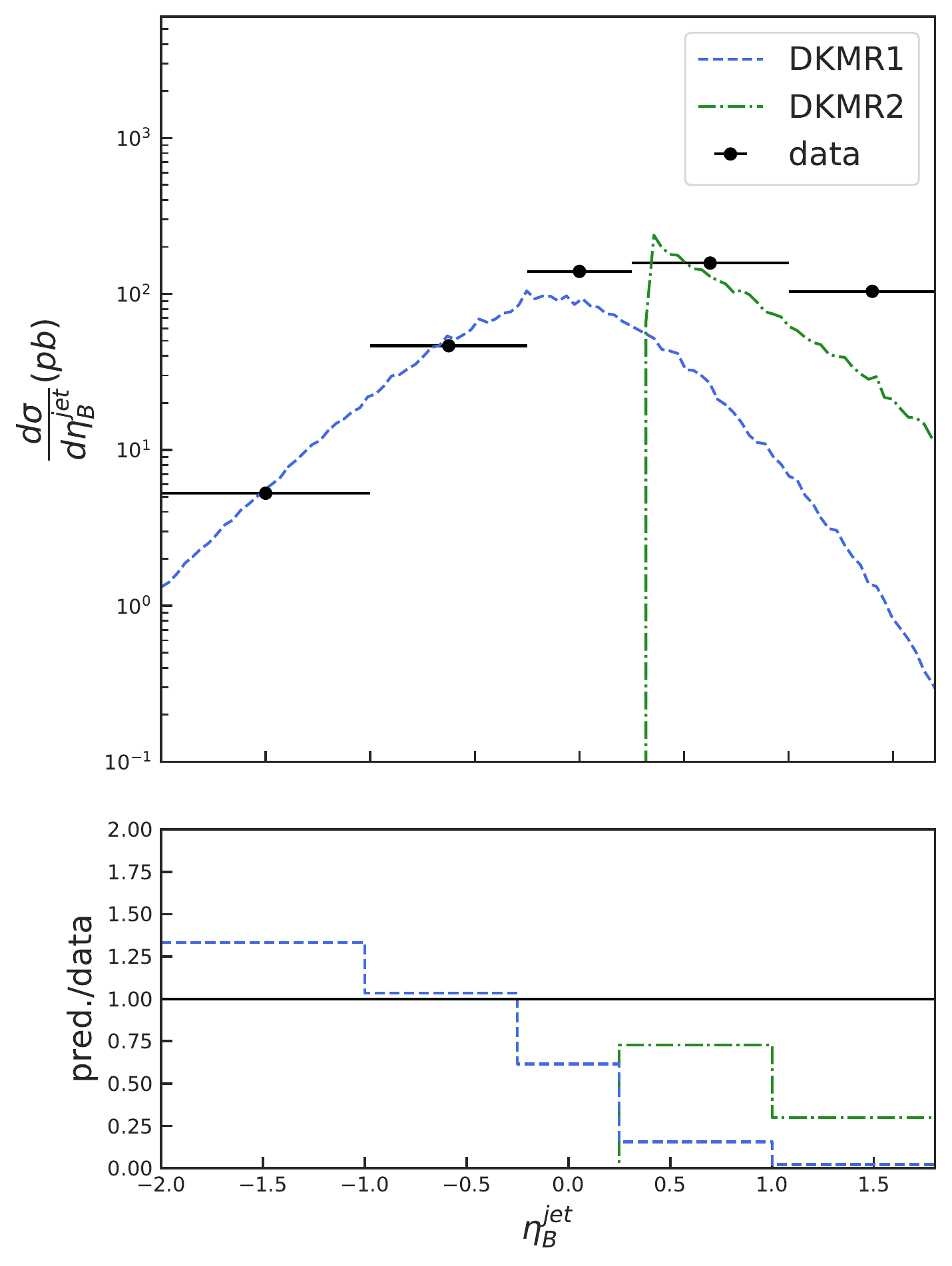}
    \includegraphics[width=5cm, height=9cm]{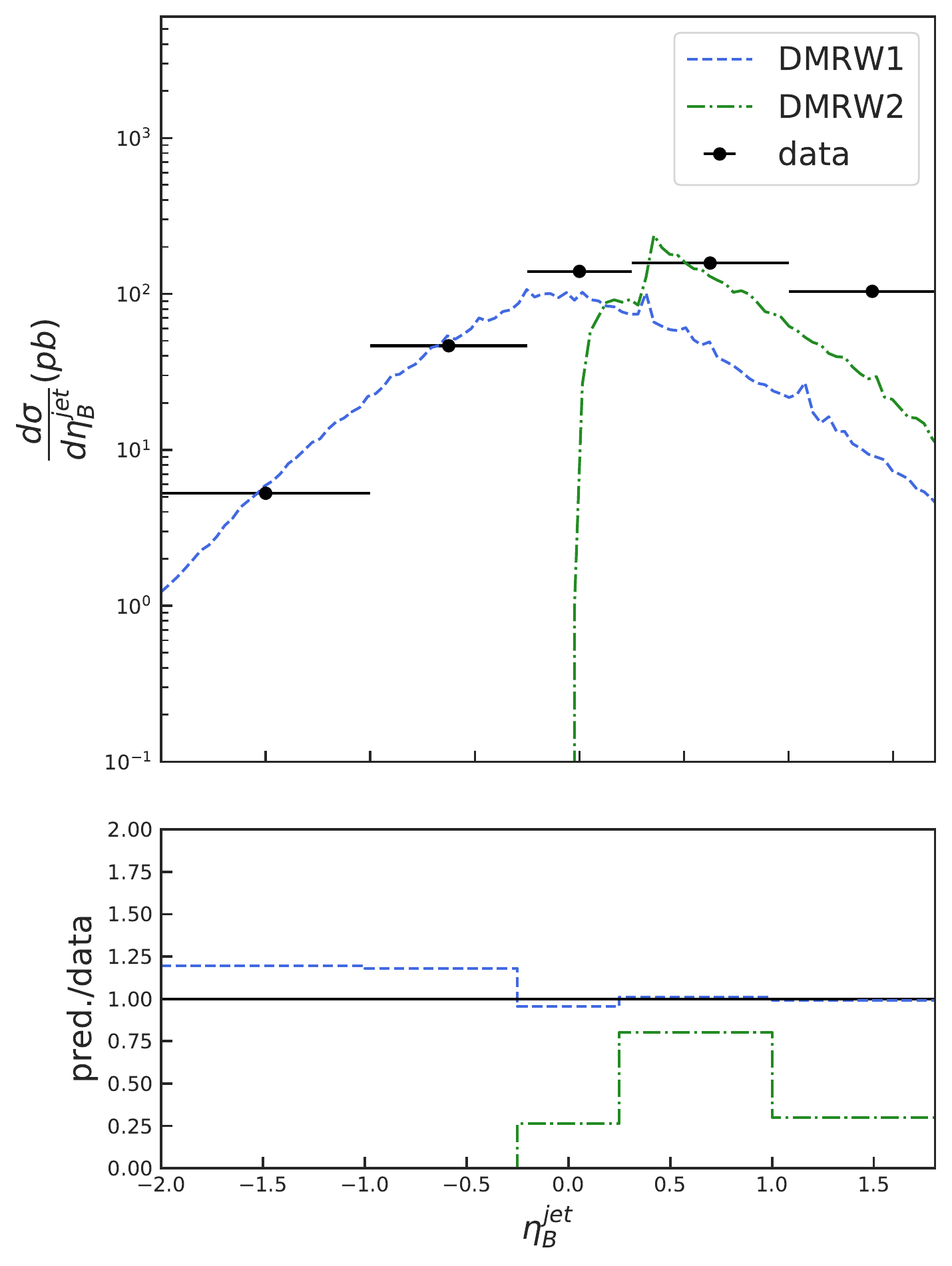}
    \includegraphics[width=5cm, height=9cm]{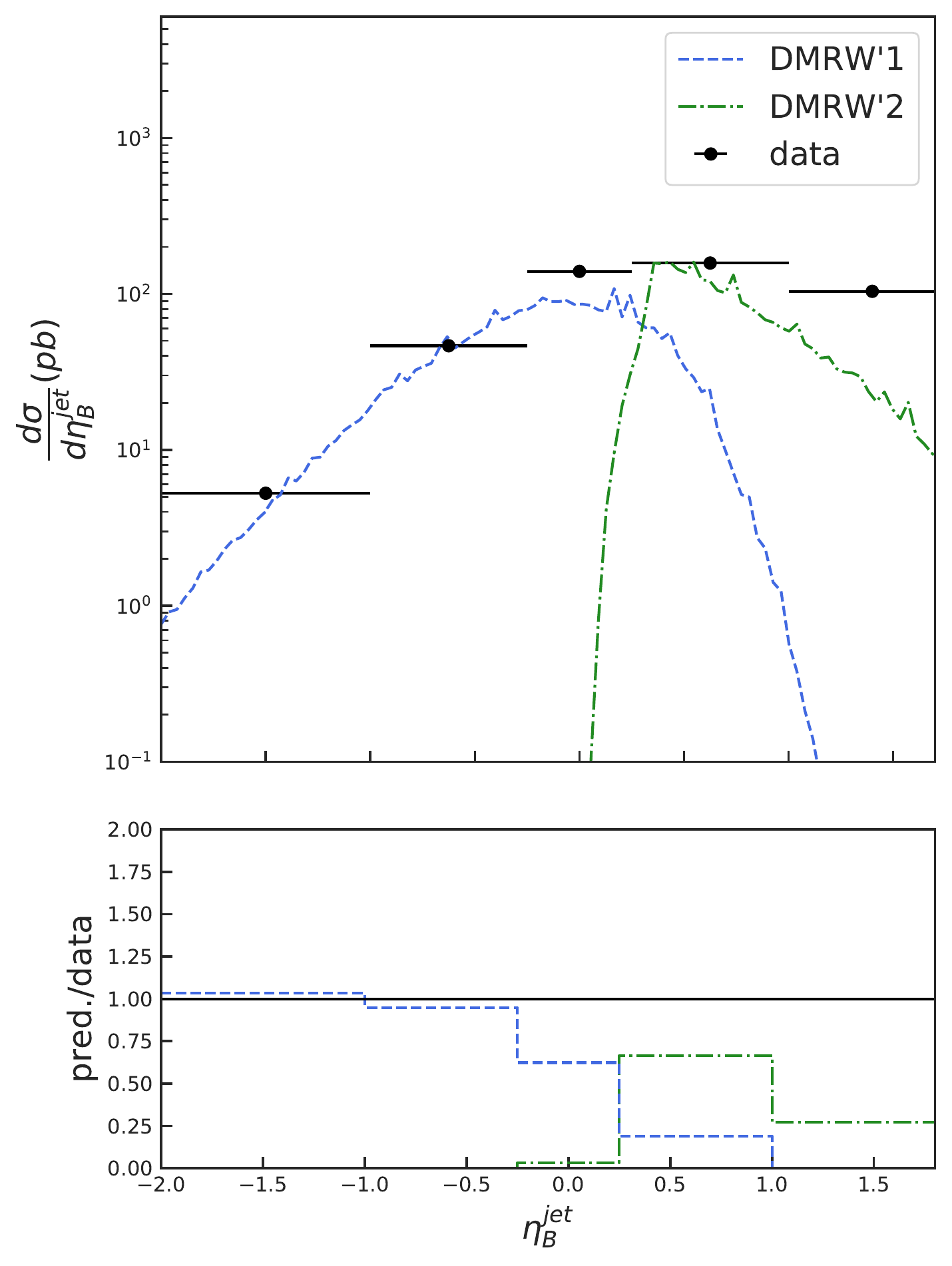}
    \caption{The top panel is the same as the figure $5$ but for $d\sigma /d\eta^{jet}_{Breit}$. The bottom panel is  $d\sigma /d\eta^{jet}_{Breit}$ but for the current (DKMR1, DMRW1, and DMRW$^\prime$1) and  the last step (DKMR2, DMRW2, and DMRW$^\prime$2)  emitted  jets for  different DUPDFs models  .
        }
   \label{fig:9}
\end{figure}

\begin{figure}
    \includegraphics[width=7cm, height=9cm]{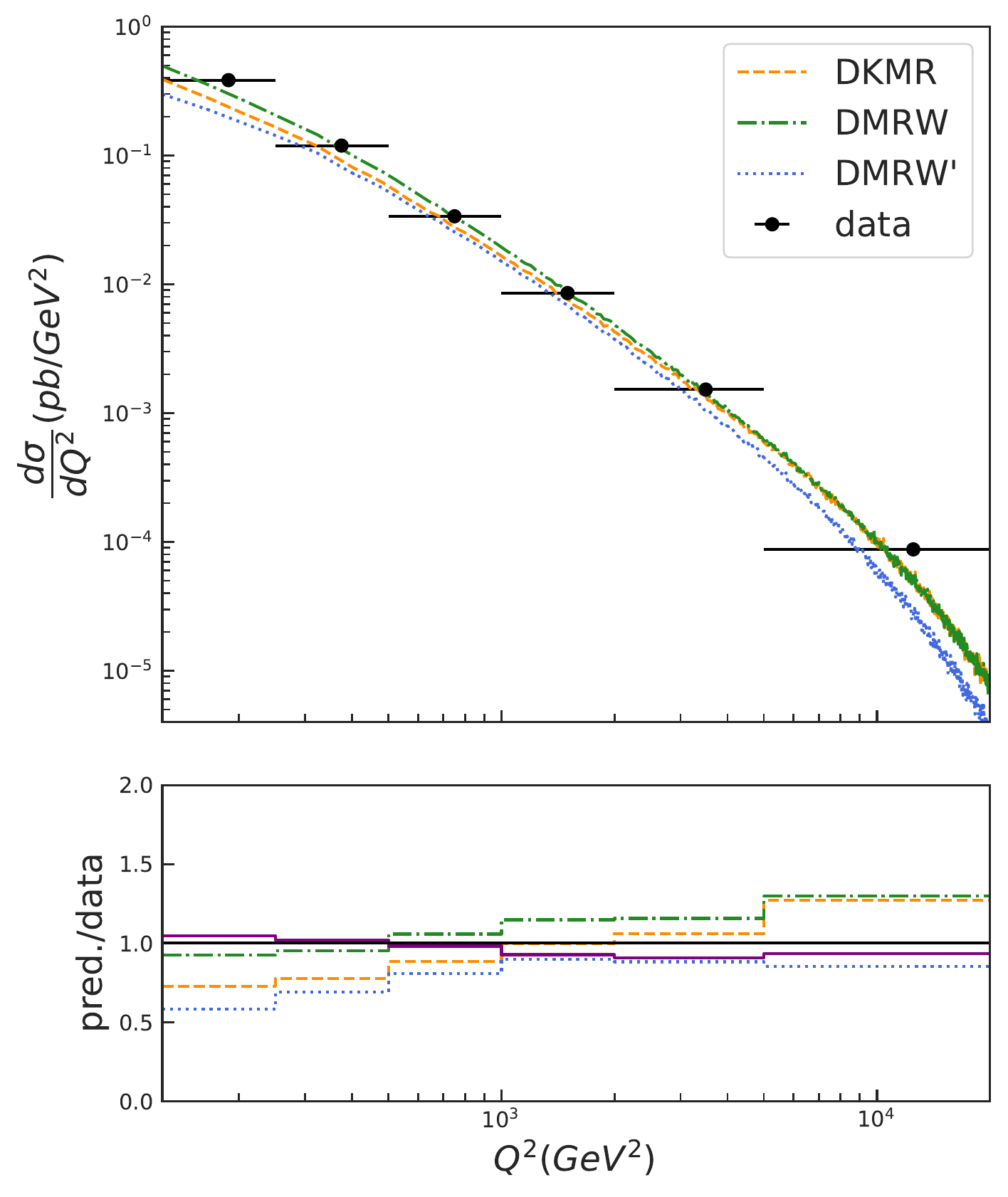}
    \includegraphics[width=7cm, height=9cm]{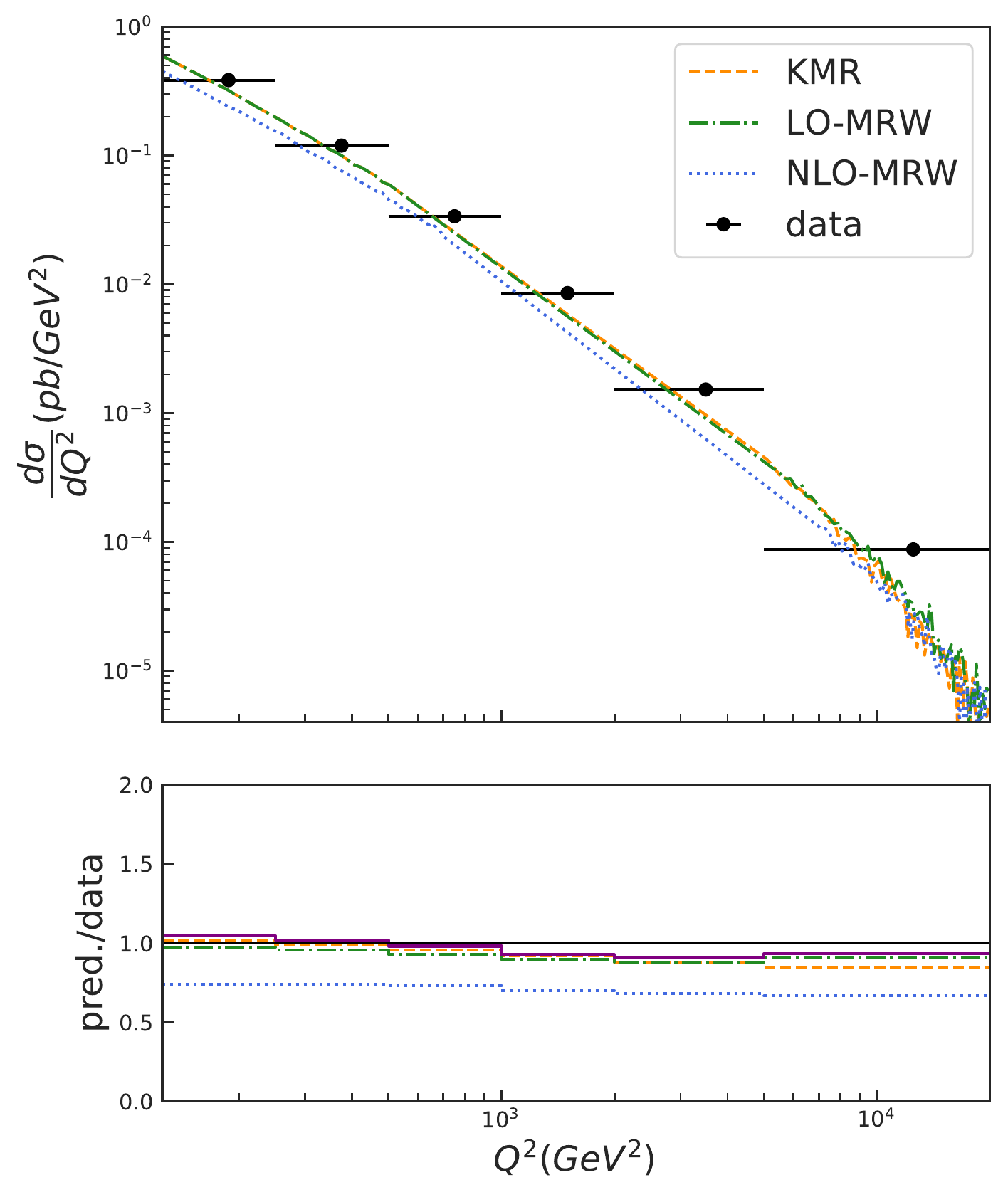}
    \caption{The same as the figure 4 but for the  inclusive dijet production. The experimental data  and the NNLO collinear result  are from the references \cite{Zeus_2010} and  \cite{nnlo_dijet}, respectively.
        }
    \label{fig:10}
\end{figure}

\begin{figure}
    \includegraphics[width=7cm, height=9cm]{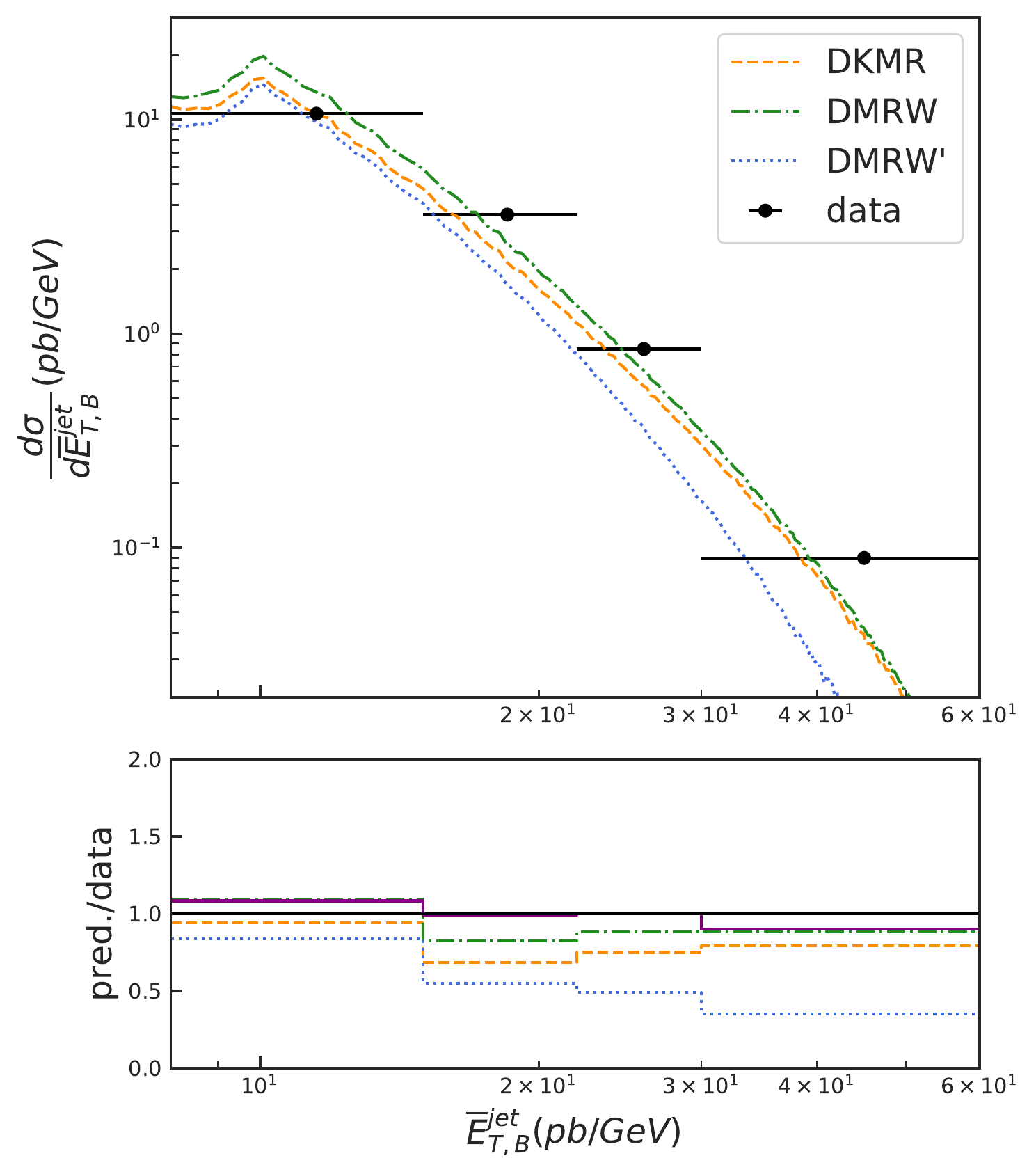}
    \includegraphics[width=7cm, height=9cm]{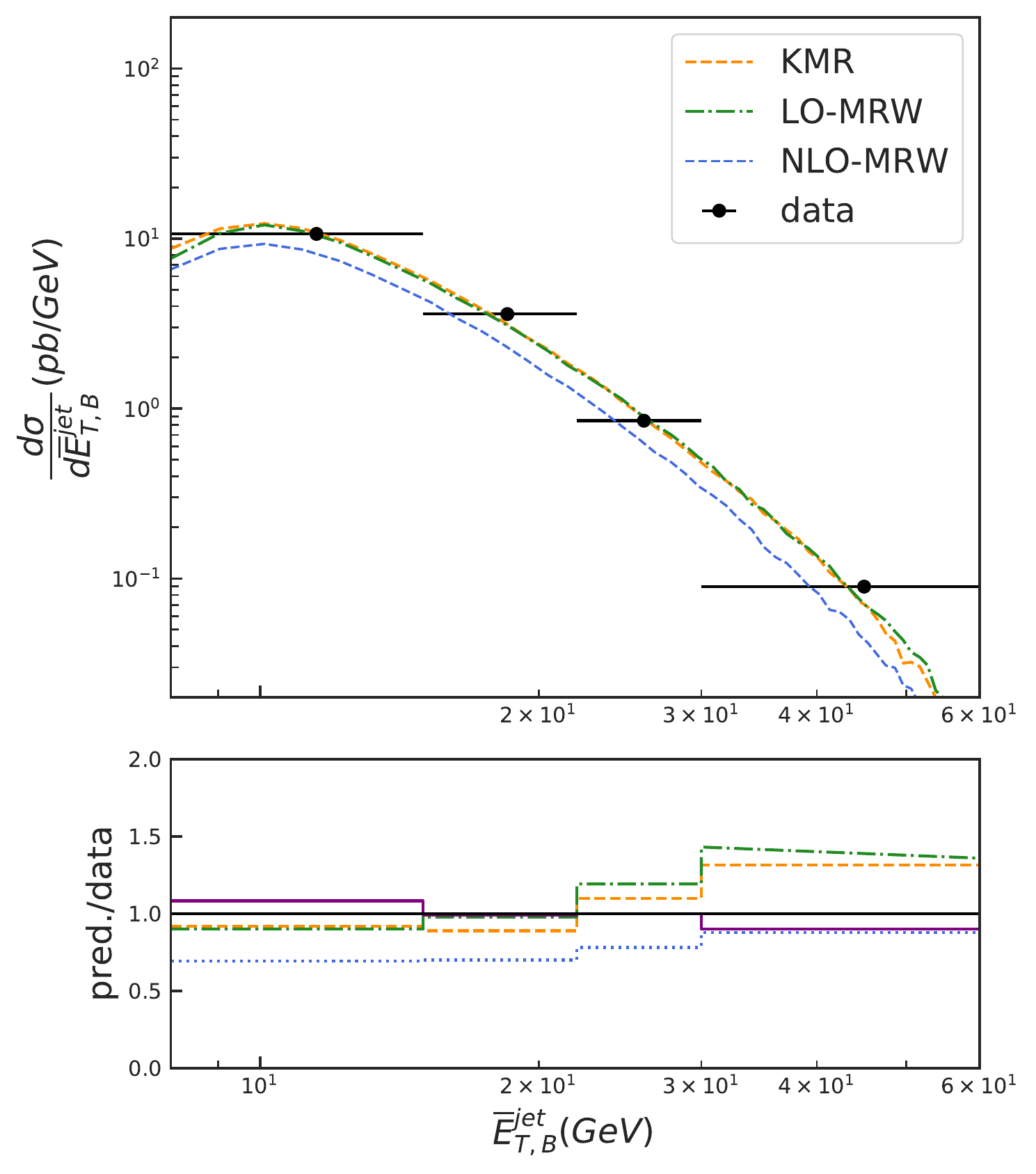}
    \caption{The same as the figure $10$ but for $d\sigma /d \overline{E}_{T,B}^{jet}$.
        }
      \label{fig:11}
\end{figure}

\begin{figure}
    \includegraphics[width=7cm, height=9cm]{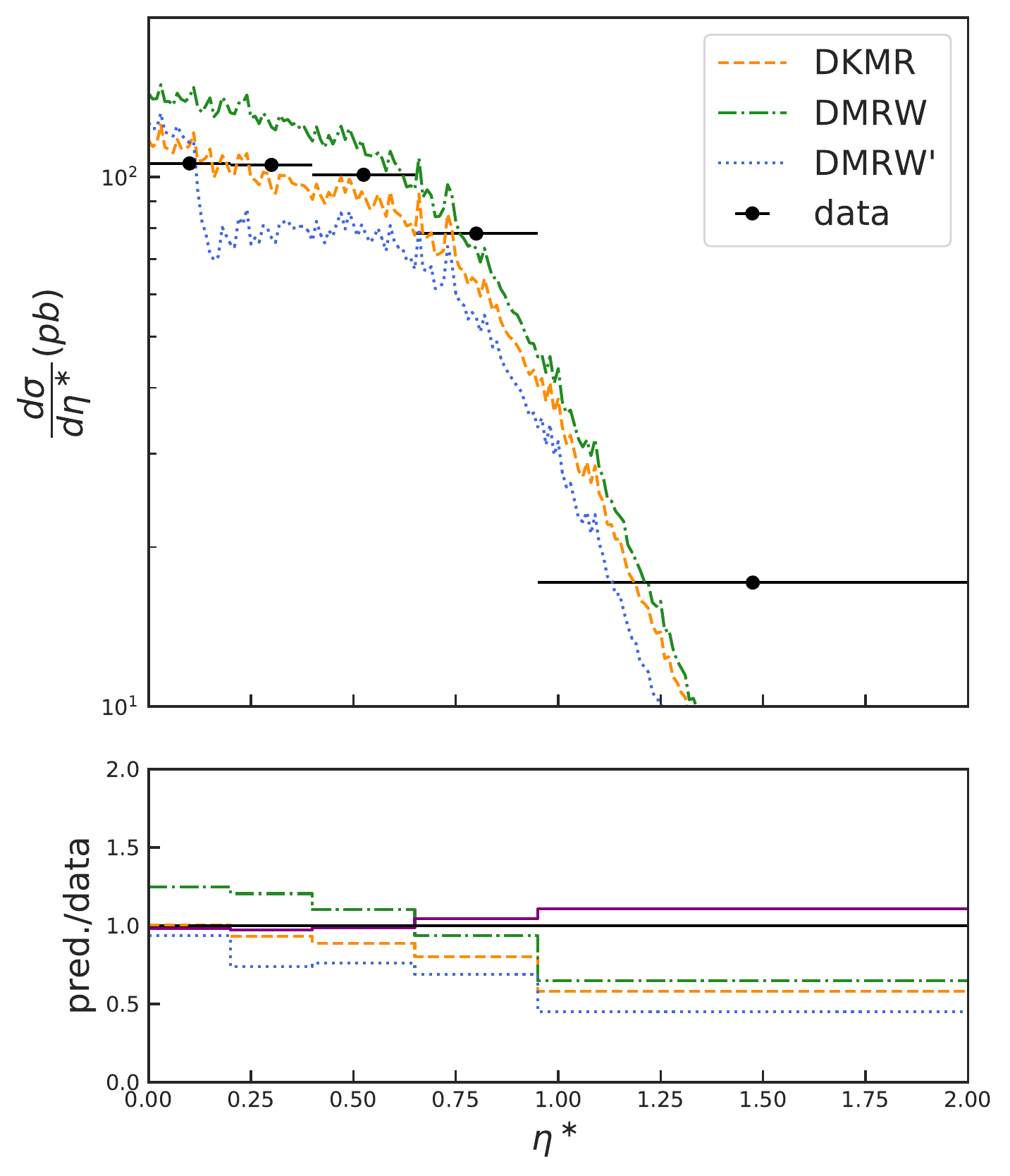}
    \includegraphics[width=7cm, height=9cm]{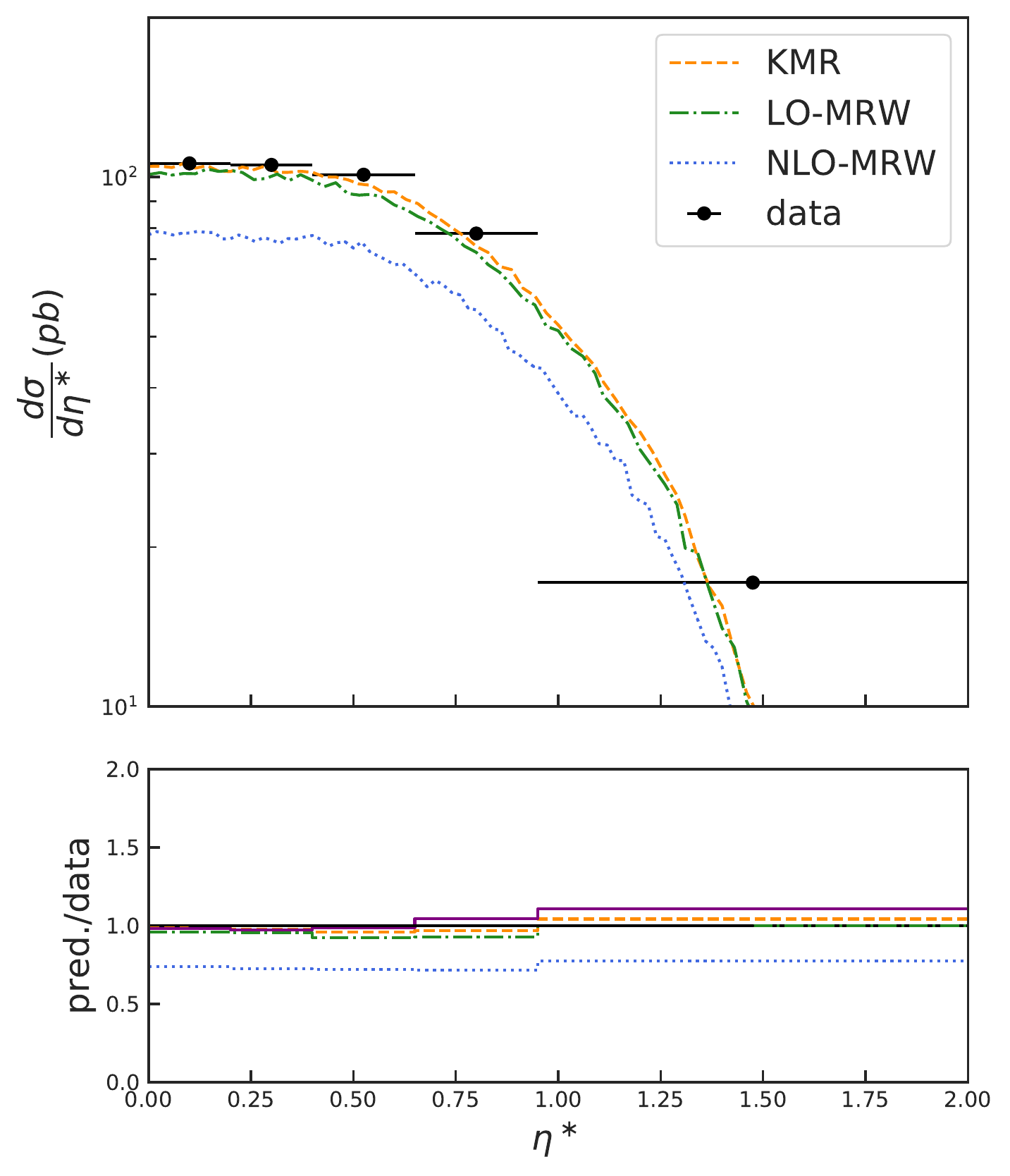}
\caption{
            The same as the figure $10$ but for $d\sigma /d\eta^\ast$.
        }
    \label{fig:12}
\end{figure}

\begin{figure}
    \includegraphics[width=7cm, height=9cm]{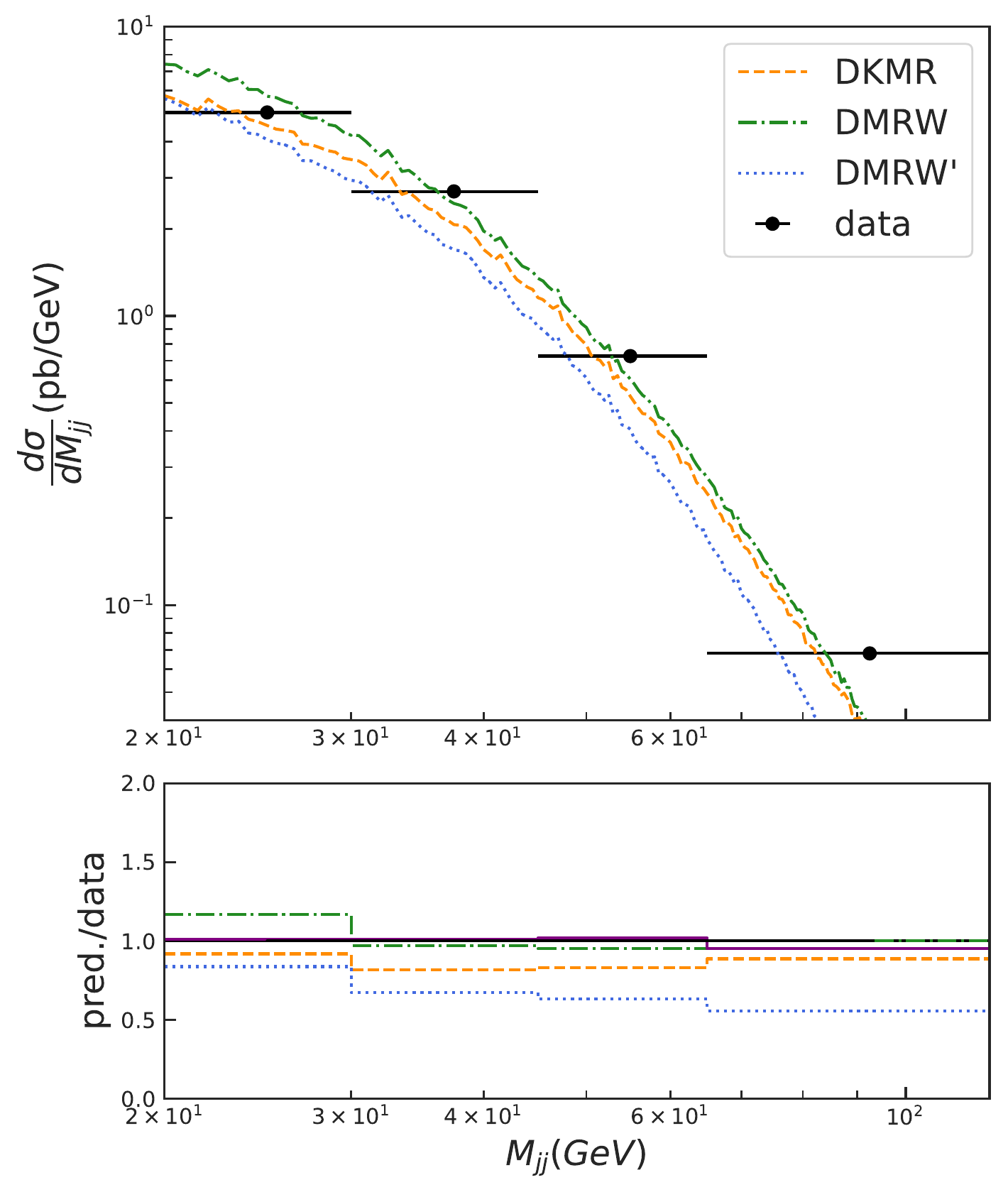}
    \includegraphics[width=7cm, height=9cm]{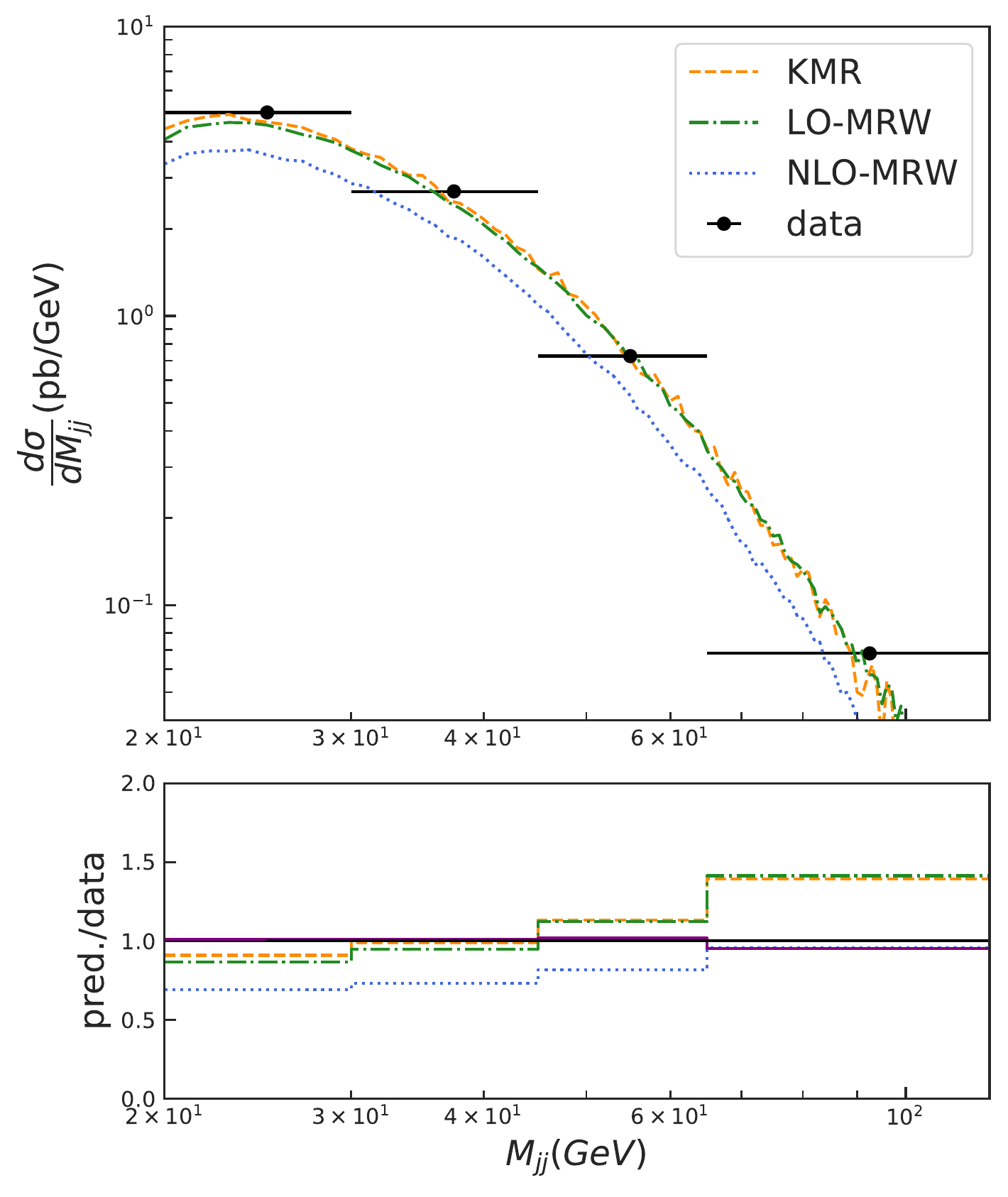}
\caption{ The same as the figure $10$ but for $d\sigma /dM_{jj}$.
        }
      \label{fig:13}
\end{figure}

\begin{figure}
    \includegraphics[width=7cm, height=9cm]{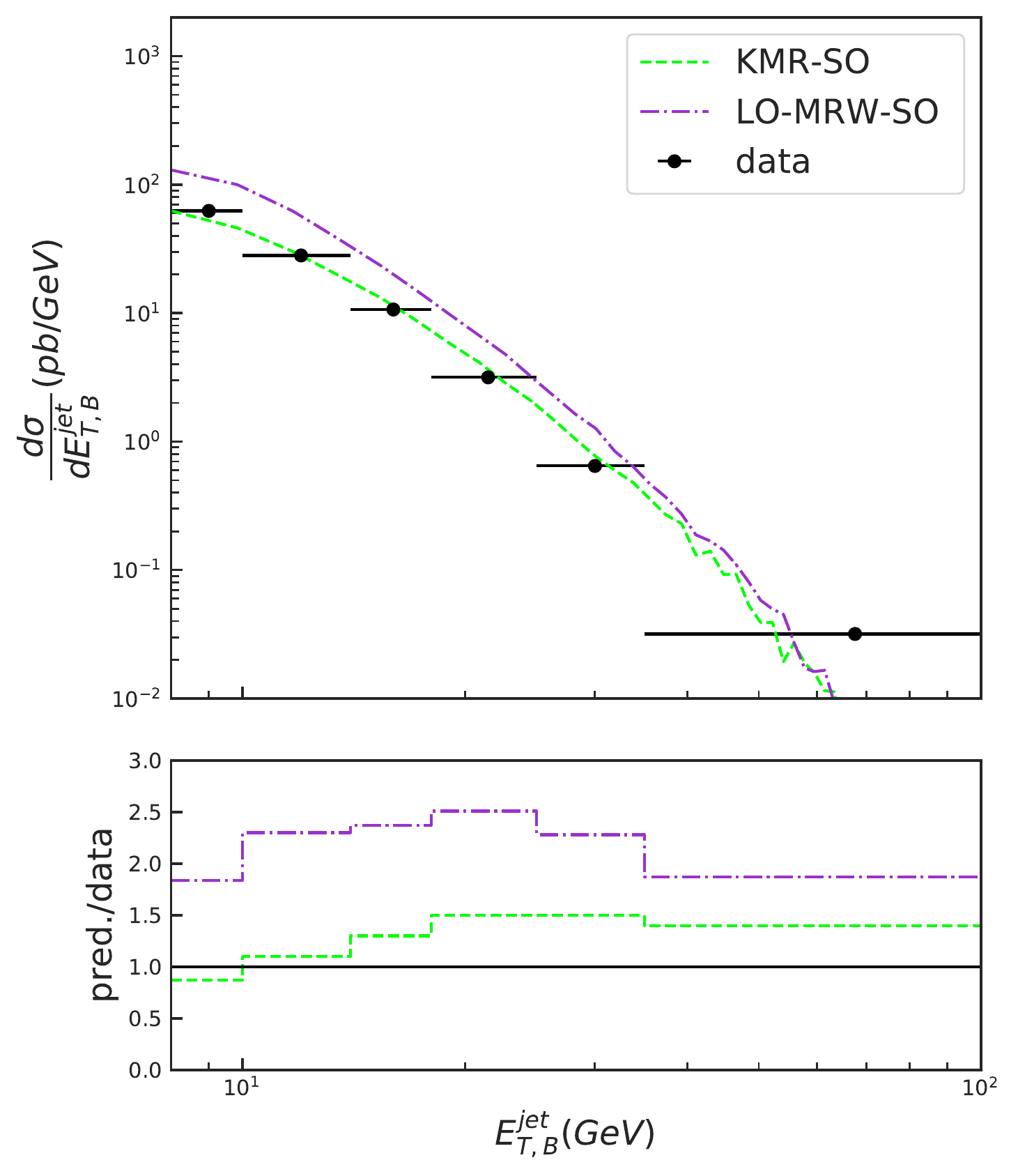}
    \includegraphics[width=7cm, height=9cm]{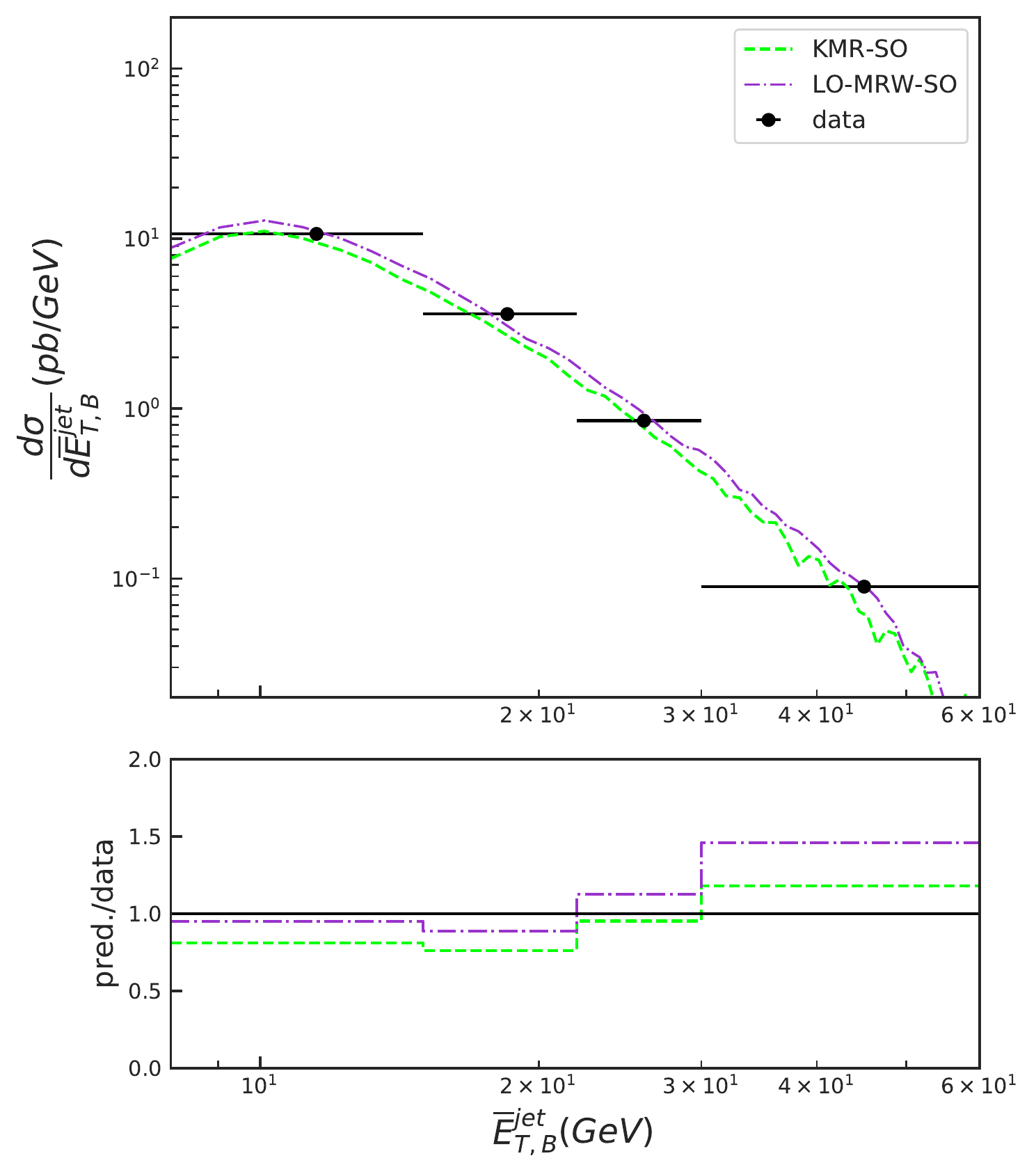}
\caption{ The same as the figure $6$ ($10$), left (right) panel, but for the KMR and  LO-MRW strong ordering constraint.} 
    \label{fig:14}
\end{figure}

  Before discussing our results, we first give an
overview of different DGLAP based UPDFs and  DUPDFs that we use
in this work as follows (these abbreviations are also used to show the related differential cross sections in the figures \ref{fig:4} to \ref{fig:14}): 

\begin{itemize}
    \item UPDFs:
    \begin{enumerate}
    \item  KMR: This formalism assumes that the parton evolves up to the last evolution step according to the DGLAP evolution equation, but in this step it gains transverse momentum and does not radiate any real emissions further more. Therefore   no real emission to the factorization scale is re-summed by the Sudakov form factor. In this method, the soft gluon emission singularity is cured by using the angular ordering of the soft gluon,  the quark and anti-quark, emissions (see the subsection \ref{KMR-Formalism}).   
 
 \item LO-MRW: This formalism is the same as the KMR, but without  additional angular ordering constraint   on the quark (anti-quark) emissions (see the subsection \ref{MRW-Formalism}) .   
     
\item  NLO-MRW: This formalism is the same as the LO-MRW, but with the virtuality $k^2=\dfrac{k_t^2}{(1-z)}$ for the scale in the DGLAP evolution equation, and   the strong ordering cutoff $\Theta(\mu^2-k^2)$   on the virtuality. For curing the divergency due to the soft gluon emission, the angular ordering of soft gluon emissions is imposed. Additionally, the NLO level PDFs and the splitting functions are used (see the subsection \ref{MRW-Formalism}).
    \end{enumerate}
    \item  DUPDFs:
    \begin{enumerate}
        \item DKMR: This formalism is the same as the KMR, but the integral over z is not performed (see the subsection \ref{KMRDUPDF}).
        \item DMRW: This formalism is the same as the LO-MRW, but the integral over z is not performed (see the subsection \ref{MRWDPDUF}).
        \item DMRW$^\prime$: This formalism is the same as DMRW, but with $k_t$ replaced by $k$ and also
adding a $\Theta(\mu^2-k^2)$ to prevent $k^2$ becomes larger than
$\mu^2$ (see the subsection \ref{MRWDPDUF}).
    \end{enumerate}
\end{itemize}
\begin{figure}
\includegraphics[width=7cm, height=8cm]{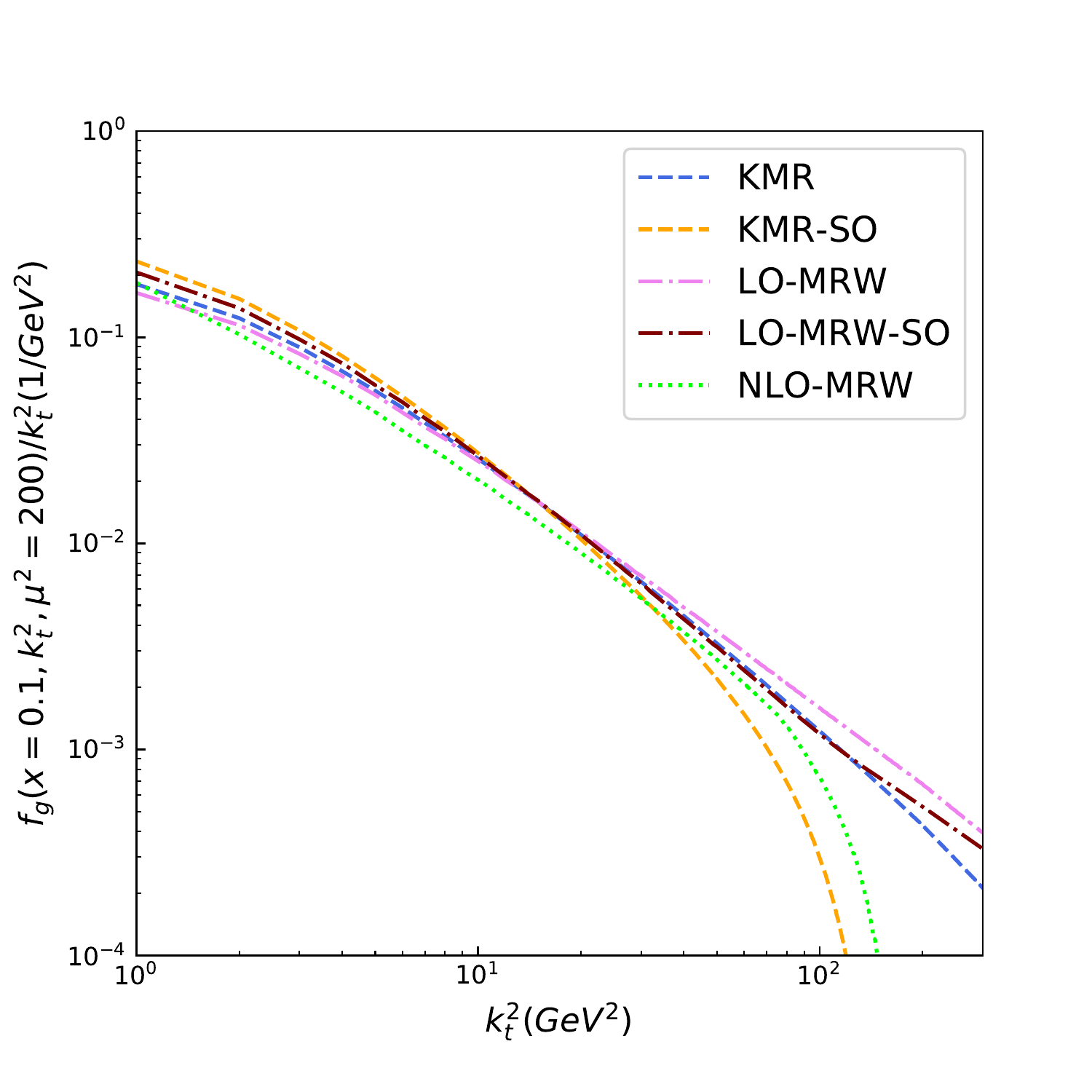}
\includegraphics[width=7cm, height=8cm]{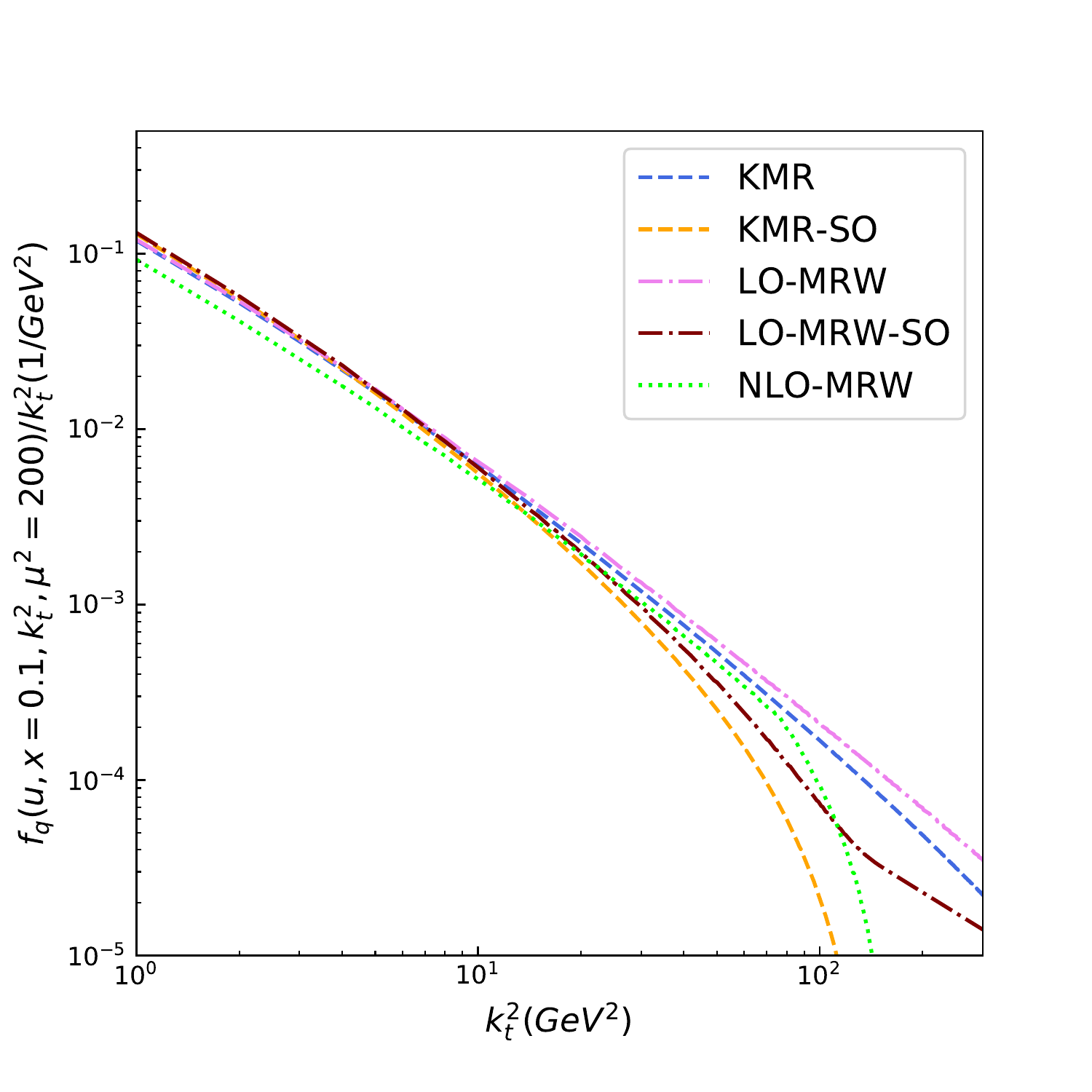}
\caption{The left and right panels show the comparison of the up quark and gluon UPDFs/$k_t^2$ of the KMR, LO-MRW and NLO-MRW (KMR-SO and LO-MRW-SO)  with the angular  (strong  ordering) cutoff, respectively. See the text for more explanations.
        }
\label{fig:3}
\end{figure}
For example, in order to gain an insight  about the behavior  of our different UPDFs, in the figure \ref{fig:3} we have plotted the KMR, LO-MRW and NLO-MRW UPDFs/$k_t^2$  for the gluon and the up quark at x=0.1 and 
$\mu^2=200$ $GeV^2$. The KMR-SO  and LO-MRW-SO are the corresponding UPDFs/$k_t^2$, but using the strong ordering constraint.  As it was discussed in the references \cite{Mod3,Mod4}, the results of different UPDFs models with the angular ordering constraint are very similar, especially at small transverse momentum. In case of strong ordering (KMR-SO), the UPDFs go to zero for $k_t^2 \sim \mu^2$. On the other hand  in the  case of MRW-SO, since the constraint is only imposed on the gluon emission terms, the corresponding distribution becomes different from the KMR-SO. We will come back to the effect of above UPDFs on the differential cross sections later on.   
 
In what follows we discuss and elaborate upon the differential cross-section predictions    
using different UPDFs, i.e., KMR, LO-MRW and NLO-MRW and DUPDFs, i.e., DKMR, DMRW and DMRW$^\prime$, in the $k_t$ and
$(z,k_t)\textrm{-factorization}s$ models, by comparing their results to
the data and to each other, see the figures  \ref{fig:4} to \ref{fig:14}. In all of these
figures, it is observed that the  LO-MRW and DMRW differential cross sections  
  are larger with
respect to  the other models presented in these figures. On the other hand    
the corresponding    KMR and LO-MRW, and DKMR and  DMRW  differential cross sections   
are close to each other.
As we pointed out before,  these similarities  can be due to  the UPDFs of the KMR
and LO-MRW which is demonstrated  at $x=0.1$ and $\mu^2=200$    in the figure \ref{fig:3}. 

The same similarity is also expected for the
corresponding differential cross sections with input DUPDFs, because only the difference between the DUPDFs
and the UPDFs are the integration over $z$ which is not performed in the
DUPDFs cases. 

Another important feature that is demonstrated in
the figure \ref{fig:3} is the smaller distribution of up quark and gluon of
the NLO-MRW  with respect to the other UPDFs except that of KMR-SO. 
Actually, this choice of virtuality for the scale, i.e.,
$k^2= \dfrac{k_t^2}{(1-z)}$, leads to the larger argument of the
PDFs with respect to the KMR and  LO-MRW UPDFs. Because we are working in the region that
the larger fractional momenta play the significant role.  Therefore the
strong ordering cutoff, $\Theta(\mu^2-k^2)$, on the virtuality in
the NLO-MRW UPDF and the DMRW$^\prime$ is a crucial factor in the sharp
decrease of these UPDFs and DUPDFs at this energy scales. These behaviors also show themselves directly in the differential cross section calculations.

 Additionally, by looking at
the bottom panels of each plot, i.e., the figures \ref{fig:4} to \ref{fig:13}, one finds that the results of the
NLO \cite{zeus2002} and NNLO \cite{nnlo_dijet} of the collinear
factorizations (denoted by solid-purple color) have better agreement to
the data with respect to our calculation. However, as it was discussed in the reference \cite{DUPDFElectroweak}, we should point 
out that, 
the result of present calculation should not be as good as
collinear approaches, on the other hand, it is more simplistic,
considering the computer time consuming, i.e.,
the fewer number of tree-level Feynman diagrams are needed for calculation in the $k_t$-factorization
framework with respect to the collinear one for calculation of the corresponding differential cross sections. Hence such feature makes the
$k_t$-factorization computationally more efficient (in order to
see more the details on how higher order diagrams are included in the tree
level lower order diagrams in the $k_t$-factorization formalisms, one can see the section $3$ of the
reference \cite{smallXReview1}). Despite the fact that this
framework is more convenient than the collinear factorization
approach, however the complications which arise by including the
transverse momentum of the evolving parton into calculation, and
obtaining an evolution equation, that can perfectly work in all
kinematic regimes, becomes a hard task. For example the
Catani–Ciafaloni-Fiorani–Marchesini (CCFM) evolution equation
\cite{CCFM1,CCFM2,CCFM3,CCFM4} which is one of the successful
evolution equations, has the limitation that cannot give
transverse momentum densities for all quark flavors and is limited
to the valence flavors of up and down quarks. Therefore, testing
different UPDFs models and obtaining a suitable approach to create
the UPDFs have significant importance in the current field of high
energy.

In the figures \ref{fig:4} to \ref{fig:9}, we compare predictions of the differential cross sections in the $k_t$ and $(z,k_t)$-factorizations with the data only for the inclusive jet production. It is observed   that the KMR and LO-MRW  predictions of  differential cross sections     overshoot the data in the figures \ref{fig:6} and \ref{fig:8}. The main factor for such a behavior   is linked to the large contribution of the Born level subprocess, see figures \ref{fig:5}, \ref{fig:7} and \ref{fig:9} in which the Born level and dijet sub-processes plotted separately. We should note here that the Born level subprocess, is not included in  the collinear factorization calculation. In fact, in the Breit frame, within the collinear factorization approach \cite{zeus2002}, the parton that comes into the hard interaction has only the component of momentum along the direction of   proton, i.e., $\hat{z}$. Therefore, the parton struck with the virtual photon which   has also a component along the $-\hat{z}$ direction, and it  scattered back with zero transverse momentum. The contribution of such process in the collinear-factorization approach is completely suppressed due to the experimental cut on $E_{T,B}^{jet}$. 

It can be visualized from the figures \ref{fig:4} and \ref{fig:6}, i.e.  $d\sigma /dQ^2$ and $d \sigma/dE_{T,B}^{jet}$, that the results of the NLO-MRW are relatively good, while it gets worse for $\eta_{B}^{jet} < 0$, in $d\sigma/d\eta_{B}^{jet}$, due to the large contribution of the Born level sub-process.  

The contribution of  Born level and  dijet sub-processes (see the sub-processes given in the figure \ref{fig:1})   are separately shown in the    figures \ref{fig:7} and \ref{fig:9}. It is shown that the results of   dijet contributions with the KMR and MRW UPDFs can only give relatively more satisfactory results with respect to the Born-level sub-process. So the overshoot of the data in the $k_t$-factorization approach, presented in the figure \ref{fig:6}, may not be connected with the double counting of the last step emission. Because in our event generation, we impose a cutoff on the transverse momentum of the incoming parton of the Born level sub-process to be less than the factorization scale, and also additional cutoff is set, which limits the transverse momentum of the incoming parton to be less than the minimum of hard jets transverse momenta. This procedure is similar to the method given in   the reference \cite{PhysRevD.100.054001}, which suppresses the hard last step emitted jets and double counting which may happen. 

In contrast to the results of the $k_t$-factorization which overshoots the data, the predictions of the
DKMR and DMRW DUPDFs of the $(z,k_t)$-factorization are in
relatively good agreement with the data of
$d\sigma/d\eta_{B}^{jet}$ and $d \sigma/dE_{T,B}^{jet}$ channels
of the inclusive jet predictions. However these predictions undershoot
the data of  $d\sigma /dQ^2$. We also see the results of the
DMRW$^\prime$ underestimates the data due to the reasons explained
before. 

One interesting point about the results of the $(z,k_t)$-factorization 
framework is that the predicted differential cross section
  have a similar behavior with respect to the dijet
in  the $k_t$-factorization method, in which one can 
observe that their  results   are in good  agreement to 
the data, except the channel of $d\sigma/dQ^2$. In other
words, we can obtain similar results  to
the dijet sub-processes in the $k_t$-factorization formalism, but just with
the inclusion of the Born level diagram.

In addition to the above points, there can be
seen a non-smooth behavior in the $\eta_{B}^{jet} \approx 0$ of
$d\sigma/d\eta_{B}^{jet}$ channel. This behavior in case of the
$(z,k_t)$-factorization differential cross section is due to the
involvement of the last step emission in the region $\eta_{B}^{jet} >
0$, i.e. there is no contribution of the last step emission in
$\eta_{B}^{jet} < 0$ region (see the predictions of different
DUPDFs models in the bottom panel of the figure \ref{fig:9}). However this
non-smooth behavior in the results  of UPDF models of the $k_t$-factorization framework (as can be seen in the top panel of the
figure \ref{fig:9}) is due to this fact that the Born level process only
contributes in the $\eta_{B}^{jet} < 0$.

The  inclusive
dijet differential cross section calculations in the $k_t$ and $(z,k_t)$-factorization approaches are
illustrated  in the  figures \ref{fig:10}-\ref{fig:13}. One can observe that the UPDF models of the KMR and
LO-MRW, and their corresponding  DUPDFs are quiet successful in
describing the data. However the results of   NLO-MRW   and
DMRW$^\prime$, due to the reasons mentioned before, underestimate the
data. It should be noted that some of our results are even
comparable with the NNLO collinear predictions, even though only
the tree level diagrams of $\gamma^\ast + q \to q + g$ and
$\gamma^\ast + g \to q + \bar{q}$ processes are calculated within the
$k_t$-factorization framework. We should also again repeat here that in
the calculation of inclusive dijet production in the $(z,k_t)$-factorization, we   only include the Born level sub-process $\gamma^\ast
+ q \to q$, while in this framework, the   $\gamma^\ast + q \to q + g$
and $\gamma^\ast +  g \to q + \bar{q}$ sub-processes   are
actually three jets processes, where the last step emission is also
directly comes into play in contrast to the $k_t$-factorization
formalism. Therefore we neglected these sub-processes in our
numerical calculation of $(z,k_t)$-factorization framework. 

It is also interesting to present the results of
$k_t$ and $(z,k_t)$-factorizations for the data of
$\dfrac{d\sigma}{dQ^2}$ in the figure \ref{fig:10}. In contrast to the
same data of inclusive jet production, the figure \ref{fig:4},  it is observe
that the results of the dijet sub-processes, using the UPDFs of
  KMR and MRW, and also their corresponding DUPDFs have  
excellent agreement with the data. 
 
Finally, considering the comparison of the results differential cross section calculation with   LO-MRW UPDF using  the  angular and strong ordering constraints,  one can clearly conclude  that considereing the    inclusive dijet data, the predictions of LO-MRW UPDFs with angular constraint does not have large discrepancy with the data. On the other hand the prediction of the LO-MRW with strong ordering cutoff is similar to the LO-MRW with angular ordering constraint. On the other hand, the prediction of the KMR with strong ordering constraint is clearly much smaller than the KMR with angular ordering cutoff. This is mostly due to additional cutoff on the non-diagonal term of the DGLAP evolution equation of the unintegrated gluon distribution function, i.e. the second term of the \cref{eq:ten}. Because in this range of energy, the quark contents of the proton have the dominant distributions, then such an additional cutoff can lead to smaller prediction of these UPDFs at this range of energy. Additionally, in the case of the inclusive jet production, the LO-MRW with the strong ordering cutoff does not have huge effect to the results  of the inclusive jet production, and its result (as can be seen in the left panel of figure \ref{fig:14}) still overshoot the data. While the result of the KMR with strong ordering cutoff, although becomes closer, but still overshoot the data. Therefore, as can be seen from the figure \ref{fig:14} using the strong ordering cutoff does not necessarily improves the predictions of the KMR and LO-MRW models, and also the predictions of the KMR and LO-MRW models do not overshoot the data significantly as suggested in the references \cite{Guiot_heavy_quark,Guiot_pathology}.
\section{Conclusions}
We calculated the inclusive jet and dijet cross sections in the
$k_t$ and $(z, k_t)\textrm{-factorizations}$ formalisms. For calculation of
the differential cross sections in the
$k_t\textrm{-factorization}$ model , we used the \textsc{KaTie} event
generator with the   $\gamma^\ast + q \to q + g$ and
$\gamma^\ast + g \to q + \bar{q}$ sub-processes, as well as $\gamma^\ast+ q \to q$ for the inclusive jet production, by considering
different  UPDF approaches, i.e., the KMR, LO-MRW
and NLO-MRW. While for the $(z,
k_t)\textrm{-factorization}$ approach we   directly
calculated   the  $\gamma^\ast+ q \to q$, and we also used the DUPDFs
analogous to the aforementioned UPDFs in the
$k_t\textrm{-factorization}$ framework.

 We observed that, in general, the differential
cross sections using the NLO-MRW UPDF underestimates the data due
to the choice of virtuality as the factorization scale of the
DGLAP evolution equation, in addition to the strong ordering
cutoff on virtuality with the hard cutoff $\Theta(\mu^2-k^2)$. It
was observed that the UPDFs of the KMR and LO-MRW formalisms show the
same behavior in this range of the center of mass energy, and the
$k_t$-factorization is inadequate in describing the data of
inclusive jet production due to the large role of the Born level
sub-process, i.e. $\gamma^\ast + q \to q$. While the same Born
level sub-process in the $(z,k_t)$-factorization was  actually the
dijet processes in the leading logarithmic approximation and gave
adequate results for all channels but not for  $d\sigma/dQ^2$.

In order to investigate the predictions made in the  references \cite{Guiot_heavy_quark,Guiot_pathology},  we also calculated $d\sigma/dE_{T,B}^{jet}$ and  $d\sigma/d\overline{E}_{T,B}^{jet}$ of the inclusive jet and dijet cross sections with the  LO-MRW and KMR UPDFs,   in which the strong ordering of the gluon emissions was used for curing soft gluon divergency. We found that this cutoff in the inclusive dijet cross section makes no significant effect on the prediction of the LO-MRW formalism, but the result of KMR with this   cutoff falls much more below the angular ordering prediction. However, in case of the experimental data of inclusive jet production, even with the strong ordering cutoff still the predictions of the KMR and LO-MRW with strong ordering constraint overshoot the data. Therefore we found that using the strong ordering cutoff does not necessary improves the predictions of the KMR and LO-MRW at these range of energies.
 
It was also found out that the $k_t$-factorization     can in fact be a suitable framework for describing the data of inclusive dijet production. Even some of its results were comparable with the   NNLO collinear factorization framework calculations. 

Finally, we observed that  due to including the last step emission into our calculation, the $(z,k_t)$-factorization can describe the data of inclusive jet and dijet cross sections relatively good, and can be a suitable alternative for the $k_t$-factorization framework.

\newpage

\end{document}